\documentclass[pra,aps,twocolumn,notitlepage,footinbib]{revtex4-2}
\usepackage{amsmath}
\usepackage{graphicx}
\usepackage{hyperref}
\usepackage{caption}
\usepackage[caption=false]{subfig}
\usepackage{color}

\begin{document}
\title{Hall effect in multi-leg bosonic ladders}
\author{E. Orignac} 
\affiliation{Univ Lyon, Ens de Lyon, CNRS, Laboratoire de Physique, F-69342 Lyon, France}
\author{R.~Citro} 
\affiliation{Physics Department ``E. R. Caianiello" and CNR-SPIN, Universitá degli Studi di Salerno, INFN, Gruppo Collegato di Salerno, 84084-Fisciano (Sa), Italy}
\author{ Thierry Giamarchi}
\affiliation{DQMP,  University  of  Geneva,  24  Quai  Ernest-Ansermet,  CH-1211  Geneva,  Switzerland} 
\date{\today}
\begin{abstract}
We use bosonization to analyze the ground state Hall response of interacting bosonic N-leg ladders threaded by a flux. We derive an explicit expression of the Hall imbalance in a perturbative expansion in the band curvature, retaining fully the interactions. For small magnetic field the Hall resistance is proportional to the derivative of the logarithm of  the charge stiffness with respect to density, generalizing the result obtained in the two leg case. We also consider the effect of temperature, and establish that at low temperature, corrections to the Hall resistance are exponentially small in the Meissner phase. 
\end{abstract}
\maketitle

\section{Introduction}
The Hall effect\cite{Hall1879} and its quantum version\cite{klitzing_IQHE} are key concepts in modern condensed matter physics, with applications in metrology\cite{Klitzing1980} and sensors\cite{ripka2009}. In the case of non-interacting electrons, the quantum Hall effect has been understood in terms of topological invariants of the ground state wavefunction\cite{thouless_1982,Avron1985,Niu1985,review_niu_2010}. For low magnetic fields, the Hall conductance obtained from the Boltzmann equation is related to the curvature of the Fermi surface by Tsuji's formula\cite{Tsuji1958a} in three dimensions, and to the area swept by the scattering path by Ong's formula\cite{Ong1991a} in two dimensions.  
However, with strongly interacting particles, the theory of the Hall effect remains challenging, even for low field\cite{Leon-PRB-2007,zotos_2000,Auerbach2018,shastry_Hall}. In the realm of
interacting quantum particles, an important issue is the Hall effect of interacting bosonic particles. Indeed in the case of bosons interactions are usually needed from the start to avoid pathological cases. A Hall effect with charged bosonic particles (Cooper pairs) can be observed in Josephson junction arrays\cite{fazio_josephson_junction_review} while a Hall effect with neutral bosonic particles has been observed  using ultracold atoms in artificial gauge field\cite{dalibard_review_artificial_gauge_fields,Galitski2013,mancini_observation_2015,dalibard_gauge_2015,Goldman2016,stuhl_ladder_2015,zhou2023strongly,impertro_strongly_2025}. The simplest interacting bosonic system in which the Hall effect can be studied is a planar ladder formed of one-dimensional interacting bosons coupled by interchain hopping, since such a system allows a nonzero flux per plaquette while being amenable to powerful analytic\cite{haldane_bosons,efetov_larkin75,cazalilla_review_bosons} and numeric\cite{white_dmrg_letter,Hallberg_rev,schollwock_DMRG_review} methods. 
The ground state phase diagram of bosonic ladders in a flux has been
considered using analytic
methods\cite{kardar_josephson_ladder,orignac01_meissner,Tokuno2014,Uchino2016a}
and numerical
methods\cite{cha2011,Piraud2014,piraud2014b,Piraud2015,DiDio2015,greschner2015,Natu2015,greschner2016,greschner2017,Orignac2017,Citro2018}. At
low flux, a Meissner-like phase, with current circulating on the edge
is found, while at higher flux, a Vortex phase-like is
obtained\cite{orignac01_meissner} that is melting above a critical
value of the interaction\cite{Orignac2017}. At finite temperature, the
crossover between the Meissner and the Vortex phase was found to
persist.\cite{Maeda_spinchain_magnetization,buser_finite-temperature_2019,Citro2018}
Experiments with ultracold atoms in artificial gauge
field\cite{dalibard_review_artificial_gauge_fields} have confirmed the
existence of Meissner and Vortex phases.\cite{atala_2014}  Moreover,
for fluxes commensurate with the particle density, analogues of the
fractional quantum Hall effect have been
predicted.\cite{Petrescu2015,Cornfeld2015,Petrescu2017,Strinati2017,strinati2019,haller_2020}
\\ 
In a previous Letter, we used a bosonization approach\cite{giamarchi_book_1d} properly taking into account band curvature \cite{schick_flux_1968,haldane_bosonisation,Leon-PRB-2007,lopatin_q1d_magnetooptical} and predicted analytically that the ground state Hall resistance of a two-leg bosonic ladder is proportional to the derivative of the logarithm of the Kohn stiffness\cite{kohn_stiffness} with respect to particle density. This result showed a remarkable connection between two important 
transport coefficients that encode the many-body effects triggered by interactions in the response to external fluxes. Recent experiments with ultracold bosons have been able to measure the Kohn stiffness \cite{schuttelkopf_drude_weight_2024}. 
Beyond the clear potential of our formula to predict Hall voltages from single chain response to Aharonov-Bohm fluxes, it clarified the exotic Hall response of charged bosonic two-leg ladder systems, in particular in the vicinity of a Mott transition\cite{donohue_commensurate_bosonic_ladder,Crepin2011}. Given the persistence of a Vortex-Meissner crossover at finite temperature\cite{buser_finite-temperature_2019}, it is also worthwhile to consider the effect of temperature on the Hall resistance in the two-leg case. 
In the present paper, we first generalize our previous ground state results to the N-leg ladder with $N\ge3$ and then calculate  in the two-leg case the temperature dependence of the Hall imbalance. In the high temperature limit, the Hall imbalance decays as a power law in temperature, while at low temperature, the ground state result is asymptotically recovered. The low temperature corrections to the ground state result are suppressed by the Boltzmann factor in the Meissner phase, while they vanish only as a power law in temperature in the Vortex phase. 

The plan of the paper is as follows. 
In Sec.~\ref{sec:three-leg}, we start by introducing the three-leg ladder model and then, using bosonization approach with an appropriate basis change, we derive an analytic expression for the Hall voltage and Hall imbalance by taking into account band curvature terms. In Sec.~\ref{sec:N-leg} we generalize the calculation to the N-leg ladder case. By introducing a basis of orthogonal polynomials in the chain index $n$, we derive  analytic expression of Hall voltage and Hall imbalance. The result shows that the proportionality of the Hall resistance  to the derivative of the logarithm of the Kohn stiffness is  present in all N-leg bosonic ladder. In Sec.~\ref{sec:thermal}, we consider the two- and three-leg bosonic ladder at positive temperature in a solvable limit, and establish that at low temperature, corrections to the Hall resistance are exponentially small in inverse temperature in the Meissner phase while they vanish as power law in the Vortex phase.

\section{Hall effect in a three-leg ladder}
\label{sec:three-leg}

We consider the Hamiltonian of a three leg ladder\cite{Kolley2015,citro2000} that in bosonized form is:
\begin{eqnarray}
H_0 &=& \sum_{j=1}^{3} \int \frac{dx}{2\pi}
\left[
u K \left(\partial_x \theta_j \right)^2
+ \frac{u}{K} \left(\partial_x \phi_j \right)^2
\right] \nonumber \\
&&-
t_\perp \sum_{j=1}^{2}  \int \frac{dx}{2\pi} \cos(\theta_{j+1}- \theta_j),
\end{eqnarray}
where the fields $\partial_x\theta_j = \pi \Pi_j$ and $\phi_j$ represent, respectively, the collective long wavelength excitations associated to the variation of the superfluid phase and the fluctuations of the density $\rho_j(x)$ of the bosons in the $j$-th chain $\rho_j(x)-\rho_0 \simeq -\frac1\pi \partial_x\phi_j(x)$ compared to the 
average density $\rho_0$. The fields satisfy the commutation relation $\lbrack \phi_j(x),\pi \Pi_{j'}(x')\rbrack=i\pi \delta_{jj'} \delta (x-x')$. The parameters $u$ and $K$ are the so-called Tomonaga-Luttinger Liquid (TLL) where $u$ is the sound velocity and $K$ a dimensionless parameter controlling the decay of the correlation functions\cite{giamarchi_book_1d,cazalilla_review_bosons,tll_review_2025}. 

We add the flux using the Landau gauge, with the vector potential along the chains, so
$A_x = B (y - y_0)$ with $y=j b$ the ordinate of chain $j$. We pick \( y_0 = 2b \), so that \( A_x = 0 \) on the central chain (\( j=2 \)), 
\(A_x = \mp B b\) on chain  $j = 1, 3$, respectively. The Hamiltonian in the presence of the flux becomes:
\begin{eqnarray}\label{eq:3chain-ham}
H &=&\sum_{j=1}^3 \int \frac{dx}{2\pi}
\left[
uK \left[\pi \Pi_j + e B b(2-j) \right]^2
+ \frac{u}{K} \sum_{j=1}^{3} \left(\partial_x \phi_j\right)^2
\right]\nonumber \\
&-& t_\perp \int \frac{dx}{2\pi} \left[
\cos(\theta_2 - \theta_1)
+ \cos(\theta_3 - \theta_2)
\right].
\end{eqnarray}
We introduce the basis transformation
\begin{equation}
\label{eq:basis}
\begin{pmatrix}
\phi_a \\
\phi_b \\
\phi_c
\end{pmatrix}
=
\begin{pmatrix}
\frac{1}{\sqrt{2}} & 0 & -\frac{1}{\sqrt{2}} \\[6pt]
\frac{1}{\sqrt{6}} & -\frac{2}{\sqrt{6}} & \frac{1}{\sqrt{6}} \\[6pt]
\frac{1}{\sqrt{3}} & \frac{1}{\sqrt{3}} & \frac{1}{\sqrt{3}}
\end{pmatrix}
\begin{pmatrix}
\phi_1 \\
\phi_2 \\
\phi_3
\end{pmatrix}, 
\end{equation}

and 
\begin{equation}
\begin{pmatrix}
\theta_a \\
\theta_b \\
\theta_c
\end{pmatrix}
=
\begin{pmatrix}
\frac{1}{\sqrt{2}} & 0 & -\frac{1}{\sqrt{2}} \\[6pt]
\frac{1}{\sqrt{6}} & -\frac{2}{\sqrt{6}} & \frac{1}{\sqrt{6}} \\[6pt]
\frac{1}{\sqrt{3}} & \frac{1}{\sqrt{3}} & \frac{1}{\sqrt{3}}
\end{pmatrix}
\begin{pmatrix}
\theta_1 \\
\theta_2 \\
\theta_3
\end{pmatrix}.
\end{equation}
In the new basis, the first term of the Hamiltonian Eq.~(\ref{eq:3chain-ham}) becomes 
\begin{eqnarray}
H&=&\int \frac{dx}{2\pi} \left[uK (\pi \Pi_a+ e B b \sqrt{2})^2 + uK (\pi \Pi_b)^2 + uK (\pi \Pi_c)^2 \right. \nonumber \\ 
&&\left. + \frac{u}{K} \sum_{\nu=a,b,c} (\partial_x \phi_\nu)^2\right],     
\end{eqnarray}
so the flux couples only to $\Pi_a$. Meanwhile, the transverse hopping term becomes 
\begin{equation}
\label{eq:hopping}
2 t_\perp \cos\!\left(\frac{\theta_a}{\sqrt{2}}\right)
     \cos\!\left(\frac{\theta_b \sqrt{3}}{\sqrt{2}}\right).
\end{equation}
In the absence of flux, such a term becomes relevant under renormalization group (RG) flow for $K>1/4$ leading to a Meissner phase\cite{Kolley2015}. When we add a flux, the term (\ref{eq:hopping}) competes with the term $2e Bb\sqrt{2}\partial_x \theta_a$ giving rise to a commensurate-incommensurate transition\cite{japaridze_cic_transition,pokrovsky_talapov_prl} and a Vortex phase\cite{Kolley2015}. Since the term \eqref{eq:hopping} gives rise to the contributions $\cos\!\left(\sqrt{6}\theta_b\right )$ and $\cos\!\left(\sqrt{2}\theta_a\right )$ under the operator product expansion (OPE), when $K>3/4$ the first term is relevant and only $\theta_a$ becomes gapless in the Vortex phase. For $K<3/4$, both $\theta_a$ and $\theta_b$ are gapless in the Vortex phase.  \\
As for the band curvature terms, \cite{schick_flux_1968,haldane_bosonisation,matveev_equilibration_2013}, they are: 
\begin{equation}
\label{eq:band-curv}
H_{bc}=
\sum_{j=1}^{3}
\alpha \, (\partial_x \phi_j)^3
+ \gamma \, (\pi \Pi_j-e A_j)^2 \, \partial_x \phi_j, 
\end{equation}
where\cite{matveev_equilibration_2013}
\begin{eqnarray}
\label{eq:alpha-def}
\alpha &=& -\frac{\partial}{\partial \rho_0} \left(\frac{u}{6\pi^2 K}\right), \\ 
\label{eq:gamma-def}
\gamma &=& -\frac{\partial}{\partial \rho_0} \left(\frac{uK}{2\pi^2}\right).
\end{eqnarray}
In the basis \eqref{eq:basis}, the first and second terms in \eqref{eq:band-curv} become

\begin{multline}
\sum_{j=1}^3 (\partial_x \phi_j)^3  = \sqrt{\frac 3 2 }(\partial_x\phi_a)^2 \partial_x \phi_b - \frac{(\partial_x \phi_b)^3}{\sqrt{6}} \\
+ \sqrt{3} \partial_x \phi_c \sum_{\nu=a,b,c} (\partial_x \phi_\nu)^2,
\end{multline}
and
\begin{widetext}

\begin{multline}
\sum_{j=1}^3 [\pi \Pi_j + eBb (2-j)]^2 \partial_x \phi_j = 2 \left(\frac{\pi \Pi_a}{\sqrt{2}} + eB \right)^2\left(\frac{\partial_x \phi_b}{\sqrt{6}} + \frac{\partial_x \phi_c}{\sqrt{3}} \right)  \\ 
 + \sqrt{8} \left(\frac{\pi \Pi_a}{\sqrt{2}} + eB \right) \partial_x \phi_a \left(\frac{\pi \Pi_b}{\sqrt{6}} + \frac{\pi \Pi_c}{\sqrt{3}}  \right) 
 +\frac{\partial_x \phi_c}{\sqrt{3}} \left[ \left(\frac{\pi \Pi_b}{\sqrt{6}} + \frac{\pi \Pi_c}{\sqrt{3}}  \right)^2 + \left(-\frac{\sqrt{2} \pi \Pi_b}{\sqrt{3}} + \frac{\pi \Pi_c}{\sqrt{3}}  \right)^2\right].   
\end{multline}
\end{widetext}
Using the expressions above in \eqref{eq:band-curv}, the relevant terms that lead to a finite Hall effect\cite{citro_hall_2025} are given by 
\begin{equation}
H'_{bc}= 2\, \gamma 
\left(
\pi \Pi_a + \sqrt{2}eB
\right)
\, \partial_x \phi_a
\left(
\frac{\pi \Pi_b}{\sqrt{6}} + \frac{\pi \Pi_c}{\sqrt{3}}
\right).
\end{equation}
In the Meissner phase, when  a current is flowing in the system along the legs,   $\left\langle\pi \Pi_c \right\rangle\ne 0$, this gives rise in the presence of flux  to a Hall polarization and Hall voltage, coming from the term $B \partial_x \phi_a \frac{\left \langle \pi \Pi_c\right\rangle}{\sqrt{3}}$. In the Vortex phase, $\langle \Pi_a\rangle\ne 0$, and the Hall polarization is reduced.   

We define the Hall polarization as 
\begin{equation}
    P_H=N_1-N_3=\int dx \lbrack \rho_1(x) -\rho_3(x)\rbrack=-\sqrt{2}\int \frac{dx}{2\pi} \partial_x \phi_a ,
\end{equation}
whose average can be calculated by perturbation theory in the band curvature term, following the procedure in Ref.~\cite{citro_hall_2025}. Using the definition of the current along the rungs $\langle j_c\rangle=e u K\sqrt{3}\langle \Pi_c \rangle$, in the first order in the band curvature term and in the Meissner phase we obtain:  
\begin{equation}
\label{eq:polarization}
 \frac{\langle P_H^{(1)}\rangle}{L} =  \frac{4}{3}\frac{\gamma B b \langle j_c \rangle}{u K}\frac{\pi^2}{2}\chi_{aa} (q=0,\omega_n=0)
\end{equation}
which is proportional to the static susceptibility,
\begin{eqnarray}
&&\chi_{aa} (q=0,\omega_n=0) = \\&&=  \frac {2}{\pi^2}\int_0^\beta {d\tau} \int_0^L dx \left\langle T_\tau \partial_x \phi_a(x,\tau) \partial_x \phi_a(0,0) \right\rangle_{H} \nonumber \\ && =\frac{2K}{\pi u},  
\end{eqnarray}
{where the subscript $H$ indicates that the expectation value is taken in the ground state of the Hamiltonian Eq.~(\ref{eq:3chain-ham}).  }
Thus we obtain a  result analogous to Eq.~(11) of Ref.\cite{citro_hall_2025} for the Hall imbalance, 
\begin{equation}
 \frac{\langle P_H^{(1)}\rangle}{B L b \langle j_c \rangle} = \frac{4 \pi}{3}\frac{ K_a}{u_c K_c u_a}\gamma,     
\end{equation} 
where we take into account explicitly possible renormalizations of the Tomonaga-Luttinger exponents and velocities for the different modes. 
Once the Hall imbalance is normalized by the flux and the applied current, it depends only on the Tomonaga-Luttinger parameters of the system. 
Besides the Hall imbalance, it is also possible to calculate the Hall resistance. Following the procedure of Ref.~\cite{citro_hall_2025}, we introduce an electrostatic potential varying linearly with the chain index to cancel the imbalance induced by the applied current, 
\begin{equation}
- e V_H (N_1-N_3) = \frac{e\sqrt{2}}{\pi} \int dx \partial_x \phi_a,    
\end{equation}
and we set $V_H$ so that the sum of the Hall imbalance and the density imbalance induced by $V_H$ cancel in first order perturbation theory. We find
\begin{equation}
V_H=-\frac{2\pi^2 B b}{3e u_c K_c} \gamma,     
\end{equation}
and injecting the {expression~(\ref{eq:gamma-def})} and noting that $u_c K_c=uK$, we end up with 
\begin{equation}
V_H=-\frac{Bb\langle j_c \rangle}{3e} \frac{\partial}{\partial \rho_0}[\ln(uK)].     
\end{equation}
We have a current density $I_x=\langle j_c\rangle/3$ now that we have 3 chains, so with $V_H=R_H B I_x$, we end up with
\begin{equation}
R_H=\frac{b}{e} \frac{\partial}{\partial \rho_0}[\ln(uK)],        
\end{equation}
where $\rho_0$ is the density without flux.
As in the two chain case \cite{citro_hall_2025}, we obtain a Hall resistance that depends on the charge of the carriers and the derivative of the logarithm of the charge stiffness with respect to particle density. Results corresponding to specific limits of our formula were obtained by Zotos et al. \cite{zotos_2000,zotos_drude_2001} from linear response theory neglecting some cross terms and by Auerbach et al. \cite{auerbach_quantum_2024} using an expansion over Krylov subspaces. In the Galilean invariant case \cite{haldane_bosons}, the Hall resistance immediately reduces to the usual expression $R_H \sim \rho_0^{-1}$. So far, we have considered a 3-leg ladder at incommensurate filling. At commensurate filling \cite{giamarchi_book_1d}, with sufficiently repulsive interaction, a Mott insulating phase forms in the chains yielding a vanishing charge stiffness. In the vicinity of the Mott insulating state \cite{schulz_cic2d}, the stiffness $uK \sim C |\rho_0 -\rho_c|$, giving a Hall resistance $R_H \sim \frac{b}{e (\rho_0 -\rho_c)}$. As in the two-chain case \cite{citro_hall_2025}, the Hall resistance diverges at the Mott transition, and its sign changes across the transition, indicating the replacement of hole-like carriers by electron-like carriers. A similar divergence is also expected near unit filling for hard core bosons \cite{citro_hall_2025}.

If we turn to the Vortex phase, as mentioned earlier, $\langle \Pi_a \rangle \ne 0$, and the Hall polarization becomes 
\begin{equation}
\frac{\langle P_H^{(1)}\rangle}{ L b \langle j_c \rangle} = \frac{ 4\pi}{3} \frac{K_a \gamma}{u_c K_c u_a} \left(B + \frac{\pi \langle \Pi_a\rangle}{e\sqrt{2}}\right),    
\end{equation}
with $\pi \langle \Pi_a \rangle \to -e \sqrt{2} B  $ at high flux. In that phase, the Hall voltage becomes proportional to $B + \pi \langle \Pi_a\rangle/ (e\sqrt{2})$ instead of $B$.

\section{Hall effect in N-leg ladders}
\label{sec:N-leg}

We now turn to the general case of an $N$-leg boson ladder with bosonized Hamiltonian
\begin{eqnarray}
    \label{eq:Nleg-bosonized}
    H&=&\sum_{n=1}^N \int \frac{dx}{2\pi} \left[u K (\pi \Pi_n -e A_n)^2 + \frac u K (\partial_x \phi_n)^2 \right] \nonumber \\ 
    && + \sum_{n=1}^N \int dx \left[\alpha (\partial_x \phi_n)^3 + \gamma (\pi \Pi_n -e A_n)^2 \partial_x \phi_n \right] \nonumber \\ 
    && - \frac{t_\perp}{\pi a_0} \sum_{n=1}^{N-1} \cos (\theta_j -\theta_{j+1}), 
\end{eqnarray}
with $[\phi_n(x),\Pi_m(y)]=i\delta_{nm} \delta(x-y)$, $u$ the velocity of excitations, $K$ the Tomonaga-Luttinger exponent. The band curvature \cite{matveev_equilibration_2013} terms are given in Eq.~(\ref{eq:alpha-def}). 
We use a Landau gauge with the vector potential $A_n=Ba (n-(N-1)/2)$. In the previous section \ref{sec:three-leg}, we have generalized the result obtained in the two-chain case \cite{citro_hall_2025} to the three chain case by finding a basis in which the flux was coupled to a single mode associated with density imbalance between the chain, while the longitudinal current was generated by an orthogonal mode and the last remaining mode remained a spectator. We now have to introduce a basis with similar properties in the case of $N$ chains: one mode carries the total current, a single orthogonal mode couples to the flux, and the remaining modes decouple.  

\subsection{Hahn polynomial basis}
To identify the observables relevant to the calculation of the Hall conductance, it is useful to introduce a basis of orthogonal polynomials in the discrete variable $n$, the chain index. Those orthogonal polynomials are known as the Chebyshev polynomials of a discrete variable \cite{erdelyi_special_functions} (p. 223 of Vol.2) , and are a particular case of the more general Hahn polynomials \cite{olver_nist_2010}.
Using the notations of \cite{olver_nist_2010}, the Hahn polynomials $Q_j(n,0,0;N-1)$ of degree $j$ satisfy the orthogonality relation 
\begin{eqnarray}
&&    \sum_{n=1}^N Q_j(n-1,0,0,N-1) Q_k(n-1,0,0,N-1) \nonumber \\ 
&& = \frac{1}{2j+1} \frac{(N+j)!(N-j-1)!}{[(N-1)!]^2} \delta_{jk}, 
\end{eqnarray}
and their explicit form is given by the Rodrigues formula \cite{olver_nist_2010} or in terms of a generalized hypergeometric function. In particular, we have 
\begin{equation}
Q_1(n,0,0,N-1)=\frac{2n+1-N}{1-N}.     
\end{equation}
For a given $N$, we define normalized polynomials $q_j(n)$ by 
\begin{equation}
q_j(n)=\sqrt{\frac{2j+1} N \prod_{l=1}^j \frac{N-l}{N+l}} Q_j(n-1,0,0,N-1),     
\end{equation}
so that 
\begin{equation}\label{eq:orthogonality-hahn}
\sum_{n=1}^N q_j(n) q_k(n) = \delta_{jk}.     
\end{equation}
This allows us to rewrite 
\begin{eqnarray}
\phi_n(x) = \sum_{j=0}^{N-1} q_j(n) \varphi_j(x), \\
\theta_n(x) =  \sum_{j=0}^{N-1} q_j(n) \vartheta_j(x), \\ 
\Pi_n(x)=   \sum_{j=0}^{N-1} q_j(n) P_j(x), 
\end{eqnarray}
with $\pi P_j = \partial_x \vartheta_j$ and $[\varphi_j(x),\vartheta_k(y)] = i \delta_{jk} \delta(x-y)$. Moreover, using the explicit expression of $q_1(n)$, we can express the vector potential 
\begin{equation}
A_n=-\frac{Ba}{2} \sqrt{\frac{N(N^2-1)} 3} q_1(n).     
\end{equation}
As shown in App.~\ref{app:hahn-lown}, for the case $N=2$, the change of variable used in \cite{citro_hall_2025} is recovered. Similarly, for the case $N=3$, the Chebyshev polynomial basis gives back the basis transformation Eq.~(\ref{eq:basis}). So the Chebyshev polynomials basis allows us to generalize the results for $N=2,3$ to general $N$. 

Without the band curvature terms, the Hamiltonian (\ref{eq:Nleg-bosonized}) is expressed as 
\begin{eqnarray}
H&=&\sum_{j=0}^{N-1} \int \frac{dx}{2\pi}\left[ u K \left(\pi P_j +\frac{eBa}2 \sqrt{\frac{N(N^2-1)}3}\delta_{j1} \right)^2 \right. \nonumber \\ 
&&  \left. +\frac{u}{K} (\partial_x \varphi_j)^2 \right] \\
&& -\frac{t_\perp}{\pi a_0} \sum_{n=1}^{N-1} \int dx \cos \left[\sum_{j=1}^{N-1} (q_j(n+1)-q_j(n)) \vartheta_j(x)\right]\nonumber, 
\end{eqnarray}
in terms of the new fields. Since the polynomial $q_0(n)$ is constant, $\vartheta_0$ does not enter interchain hopping. The polynomial $q_1$ being linear, the coefficient of $\vartheta_1$ inside the cosine is independent of $n$. In general, using Eq.~(18.22.19) of Ref.~\cite{olver_nist_2010}, the finite differences $(q_j(n+1)-q_j(n))$ can be expressed in terms of the Hahn polynomials $Q_{j-1}(n-1,1,1,N-2)$ of degree $j-1$ and orthogonal with respect to the weight $n(N-n)$. We see that the magnetic field couples only to $\vartheta_1$. This generalizes the method in the case of two chains \cite{citro_hall_2025} and three chains (Sec.~\ref{sec:three-leg}), of identifying a single phase field coupled to the flux. At low flux, a Meissner-like phase is formed \cite{orignac01_meissner} and beyond a threshold in flux a vortex-like phase is obtained. Operator products expansions give contributions {$\propto \cos (\theta_{n+1}+\theta_{n-1}-2\theta_n)$} of scaling dimension $3/(2K)$ that depend only on the fields $\vartheta_{j}$ with $j\ge 2$. For $K>3/4$, these contributions are relevant. Using twice Eq.~(18.22.19), we can express $\theta_{n+1}+\theta_{n-1}-2\theta_n$ in terms of the Hahn polynomials $Q_{j}(n-1,2,2,N-2)$. Using the orthogonality of the polynomials, we see that the minimization of the expectation value of the operators {$\cos (\theta_{n+1}+\theta_{n-1}-2\theta_n)$} gives unique values to $\langle \vartheta_{j}\rangle$ (up to periodicity), so that in the Vortex phase, only $\vartheta_0$ and $\vartheta_1$ are gapless if $K>3/4$. This is a direct generalization of the results in Sec.~\ref{sec:three-leg}
\subsection{Calculation of the Hall resistance}
To obtain the Hall imbalance, we need to consider the band curvature terms in the new basis. More precisely, we need to consider the term that couples the total current $\propto P_0$ to the magnetic flux $B$. Such a term can only come from $\gamma \sum_n (\pi \Pi_n -e A_n)^2 \partial_x \phi_n$. 
It has to be proportional to $P_0 ({\pi} P_1+ eBa (N (N^2-1)/12)^{1/2}) \partial_x \varphi_j$, and the orthogonality relations, Eq.~(\ref{eq:orthogonality-hahn}),  then impose $j=1$.  
So the term giving rise to the Hall effect in the Meissner phase is 
\begin{equation}
    \gamma e B a \sqrt{(N^2-1)/3} \pi P_0 \partial_x \varphi_1, 
\end{equation}
making the Hall imbalance a linear function of chain index. 
Repeating the calculation in \cite{citro_hall_2025} with a total current 
\begin{equation}
    \langle j \rangle = e \sqrt{N} u K \langle P_0 \rangle,  
\end{equation}
gives rise , in the Meissner phase, to a potential 
\begin{equation}
\gamma \pi e B a \sqrt{\frac{N^2-1}{3N}} \frac{\langle j \rangle}{e u K} \partial_x\varphi_1.     
\end{equation}
Such potential yields a nonzero Hall imbalance, determined by {$\langle \partial_x \varphi_1\rangle$}. We obtain a change of density on the $n$-th chain
\begin{equation}
-\frac 1 \pi \langle \partial_x \phi_n \rangle = \frac{2\pi \gamma Bb K_1}{u_0 u_1 K_0} \left(n -\frac{N+1}{2}\right),    
\end{equation}
that is an affine function of the chain index. Defining the Hall imbalance as $\langle P_H \rangle = N_N-N_1$, we find
\begin{equation}
\frac{\langle P_H \rangle} L = \frac{2\pi \gamma B a K_1}{u_0 u_1 K_0}.     
\end{equation}
The Hall imbalance is nonzero, unless the potential generated by the combination of current and flux is compensated by an applied external potential 
\begin{eqnarray}
 V_H &=& - e E a \sum_{j=0}^{N-1} \left(j -\frac{N-1}{2}\right) \frac{\partial_x \varphi_j}{\pi} \nonumber \\ 
 &=&\frac{eE a} {2\pi} \sqrt{\frac{N(N^2-1)}{3}} \partial_x \varphi_1.  
\end{eqnarray}
The cancellation gives the value of the electric field $E$ along the 
rung direction
\begin{equation}
E=-\frac{B}{e} \frac{\langle j \rangle} N \frac{\partial}{\partial \rho_0} \left[\ln (u K) \right],     
\end{equation}
which generalizes the result of \cite{citro_hall_2025} to the case of the $N$-leg ladder. If we note that the current density is $I_x=j/N$, we find a Hall resistance identical to the 3 leg case. As in the two-chain and three-chain cases, with a Galilean invariant system, we recover the usual formula for the Hall resistance at low flux. Also, we recover the sign change and divergence of Hall resistance across a Mott transition.  

An alternative method to obtain the Hall resistance at high frequency has been proposed in \cite{Shastry1993}. The Hall resistance is given by 
\begin{equation}
    R_H^* = \lim_{B\to 0}\frac{L}{iB} \frac{\langle [J_\parallel,J_\perp] \rangle} {\langle \frac{\partial^2 H}{\partial A_\parallel^2 } \rangle \langle \frac{\partial^2 H}{\partial A_\perp^2 } \rangle }, 
\end{equation}
{where $J_\parallel$ is the total current along the chains, $J_\perp$ is the total current along the rungs, $A_\parallel$ is the projection of the vector potential along the chain direction, $A_\perp$ its projection along the rungs.}  
Using the full Hamiltonian, we have 
\begin{eqnarray}
J_\parallel = \sum_{j=1}^N e u K \left(\Pi_j - \frac{eA_j}{\pi} \right) -2e \gamma \left(\pi \Pi_j - eA_j \right) \partial_x \phi_j,     
\end{eqnarray}
\begin{eqnarray}
J_\perp = \frac{2t_\perp e a}{\pi a_0} \sum_{j=1}^{N-1} \sin \left(\theta_{j}-\theta_{j+1}\right),     
\end{eqnarray}
\begin{eqnarray}
 \langle \frac{\partial^2 H}{\partial A_\parallel^2} \rangle= -NL e^2\frac{uK}\pi,   
\end{eqnarray}
and 
\begin{eqnarray}
 \left\langle \frac{\partial^2 H}{\partial A_\perp^2} \right\rangle= -L\frac{2t_\perp e^2 a^2}{\pi a_0} \sum_{j=1}^{N-1} \langle \cos \left(\theta_{j}-\theta_{j+1}\right)\rangle.   
\end{eqnarray}
The expectation value of the commutator is 
\begin{eqnarray}
\langle [J_\parallel,J_\perp] \rangle =\frac{4it_\perp e^3 a^2} {a_0} B \gamma L \sum_{j=1}^{N-1} \langle \cos \left(\theta_{j}-\theta_{j+1}\right)\rangle,     
\end{eqnarray}
so we obtain 
\begin{equation}
 R_H^*= -\frac 1 {Ne} \frac{\partial [\ln (uK)]}{\partial \rho_0}.    
\end{equation}
The expression of the high frequency Hall resistivity matches the one we derived for low frequency using a perturbation theory at first order in the band curvature. This hints that the Hall resistivity of the bosonic ladder system has a weak dependence of current frequency. 

\section{Effect of temperature on Hall imbalance}\label{sec:thermal}
\subsection{Two-leg lader}
\label{ssec:two-leg}
\begin{figure}
    \centering
    \includegraphics[width=\linewidth]{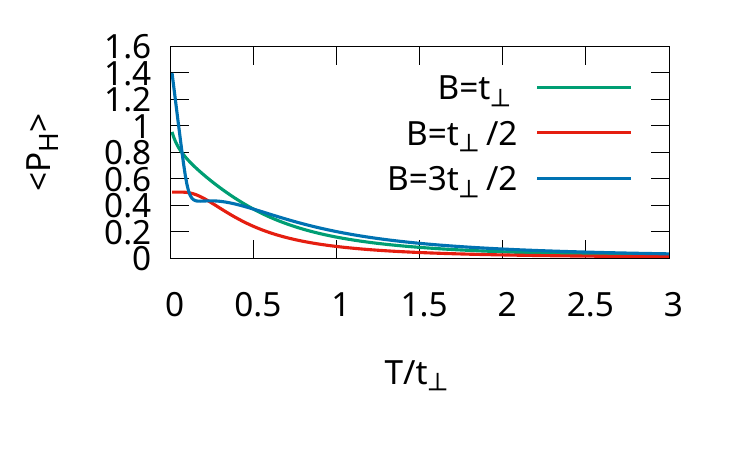}
    \caption{Temperature dependence of the Hall imbalance in the Meissner phase (Red) in the Vortex phase (Blue) and at the commensurate-incommensurate transition (Green). {In the Meissner phase a near plateau is seen for $T<0.1 t_\perp$. By contrast, in the Vortex phase, and at the transition, the Hall imbalance drops rapidly as temperature increases when $T<0.1 t_\perp$. In all the phases, for $T\gg t_\perp$, the Hall imbalance decays as $1/T^2$.  } }
    \label{fig:LE_imbalance}
\end{figure}

In Ref.~\cite{citro_hall_2025}, we derived a general expression of the Hall imbalance in the two chain case, but we considered only the zero temperature case. Temperature is expected to blur the sharp features of the Meissner-vortex transition of the ladder\cite{Citro2018}. It is thus necessary to ascertain that a nonzero temperature still yields a nonzero Hall imbalance when both longitudinal current and artificial gauge field are present. We will assume non-zero Kohn stiffness in the isolated chains when the temperature is strictly positive. This can be realized when the chains are described by an integrable model such as the Lieb-Liniger model\cite{lieb_bosons_1D,amico04_boson_integrable_review}.    The expression~\cite{citro_hall_2025} 
\begin{widetext}
\begin{eqnarray}\label{eq:imbalance-prl}
\frac{\langle P_H \rangle}{\langle j_c \rangle} &=& \frac{\gamma \sqrt{2}}{e u_c K_c} \int_0^\beta d\tau \int dx_1 \int dx_2 \left\langle T_\tau \left[(\pi \Pi_a - a A_a) \partial_{x_1} \phi_a\right](x_1,\tau)  \partial_{x_2} \phi_a  (x_2,\tau) \right\rangle,  
\end{eqnarray}
\end{widetext}
can be evaluated for $\beta<+\infty$ at the Luther-Emery point \cite{luther_emery_backscattering}, $K_a=1/2$. 
For that, we first use the rescaling $\theta_a=\sqrt{2}\theta$, $\Pi_a=\sqrt{2}\Pi$ and $\phi_a=\phi/\sqrt{2}$, and rewrite 
\begin{eqnarray}\label{eq:sine-gordon}
H_a=\int \frac{dx}{2\pi} u_a \left[\left(\pi \Pi -\frac{eBa}{2}\right)^2 +(\partial_x\phi)^2 \right]-\frac{t_\perp}{\pi a_0} \int dx \cos 2\theta. \nonumber \\       
\end{eqnarray}
We then introduce the fermion fields
\begin{eqnarray}
\psi_R(x)=\frac{e^{i(\theta-\phi)(x)}}{\sqrt{2\pi a_0}} \\ 
\psi_L(x)=\frac{e^{i(\theta+\phi)(x)}}{\sqrt{2\pi a_0}} 
\end{eqnarray}
and rewrite 
\begin{eqnarray}
H_a = \int dx \Psi^\dagger \left[-i u_a \sigma_3 \partial_x -t_\perp \sigma_1 -u_a \frac{eBa}2 \openone\right] \Psi,      
\end{eqnarray}
where 
\begin{equation}
    \Psi = \left(\begin{array}{c} \psi_R \\ \psi_L^\dagger \end{array} \right),
\end{equation}
and $\sigma_{1,2,3}$ are Pauli matrices \cite{landau_mecaq}. When $B=0$, $H_a$ is the Hamiltonian of non-interacting Dirac fermions of mass $t_\perp$. The applied flux $B$ plays the role of a chemical potential. 
Then, 
\begin{eqnarray}
&&\int dx \sqrt{2} \partial_x \phi_a = -\pi \int dx \Psi^\dagger \sigma_3 \Psi, \\ 
&&\int dx [\pi \Pi_a - e A_a ] \partial_x \phi_a = \pi \int dx \Psi^\dagger \left(i \partial_x \openone + \frac{eBa} 2 \sigma_3\right) \Psi, \nonumber \\
\end{eqnarray}
so that 
\begin{widetext}
\begin{eqnarray}
\frac{\langle P_H \rangle}{\langle j_c \rangle} = - \frac{\pi^2 \gamma} {e u_c K_c} \int_0^\beta d\tau \int dx_1 \int dx_2 
\left\langle T_\tau  \Psi^\dagger(x_1,\tau) \left(i \partial_{x_1} \openone + \frac{eBa} 2 \sigma_3\right) \Psi(x_1,\tau) \Psi^\dagger (x_2,0) \sigma_3 \Psi (x_2,0)   \right\rangle.  
\end{eqnarray}
\end{widetext}
The Hall imbalance can the be evaluated using Green's function techniques \cite{mahan_book}.
Going to Fourier space,
\begin{equation}
\psi_\nu(x) =\frac 1 {\sqrt{L}} \sum_k c_{k,\nu} e^{i k x},    
\end{equation}
we define the Nambu correlator
\begin{equation}
G(k,\tau)=-\left(\begin{array}{cc}
    \langle T_\tau c_{kR}(\tau) c^\dagger_{kR}(0)\rangle &   \langle T_\tau c_{kR}(\tau) c_{-kL}(0)\rangle\\
     \langle T_\tau c_{-kL}^\dagger(\tau) c^\dagger_{kR}(0)\rangle & \langle T_\tau c^\dagger_{-kL}(\tau) c_{-kL}(0)\rangle 
\end{array}\right),     
\end{equation}
and its Fourier transform 
\begin{equation}
G(k,\tau)= \frac 1 \beta \sum_{i\nu_n} G(k,i\nu_n) e^{-i \nu_n \tau },  \end{equation}
with $\nu_n =\pi (2n+1)/\beta$. We have 
\begin{eqnarray}
\frac{\langle P_H \rangle}{L \langle j_c \rangle} &=& - \frac{\pi^2 \gamma} {e u_c K_c} \frac 1 \beta \sum_{i\nu_n} \int \frac{dk}{2\pi} \mathrm{Tr}\left[\left(\frac{eBa}2 \sigma_3 - k \openone \right) G(k,i\nu_n) \right. \nonumber \\ 
&& \left.\sigma_3 G(k,i\nu_n) \right].
\end{eqnarray}
 We split $\langle P_H \rangle = \langle P_H^{(1)} \rangle + \langle P_H^{(2)} \rangle$ where 
\begin{eqnarray}
&&\frac{\langle P_H^{(1)} \rangle} {L \langle j_c \rangle} =   \frac{\pi^2 \gamma} {e u_c K_c} \frac 1 \beta \sum_{i\nu_n} \int \frac{dk}{2\pi} k \mathrm{Tr}\left[ G(k,i\nu_n) \sigma_3 G(k,i\nu_n) \right], \nonumber \\
\\
&&\frac{\langle P_H^{(2)} \rangle} {L \langle j_c \rangle} =  - \frac{\pi^2 \gamma B a } {2 u_c K_c} \frac 1 \beta \sum_{i\nu_n} \int \frac{dk}{2\pi} \mathrm{Tr}\left[\sigma_3 G(k,i\nu_n) \sigma_3 G(k,i\nu_n) \right],  \nonumber\\
\end{eqnarray}
so that the Hall imbalance is given by the sum of two bubble diagrams.
To evaluate these expressions, we use 
\begin{eqnarray}
G(k,i\nu_n) = \frac 1 2 \sum_{r=\pm} \left(1+r \frac{u_a k\sigma_3 -t_\perp \sigma_1}{\epsilon(k)} \right) \frac{1}{i\nu_n + h - r \epsilon(k)},      
\end{eqnarray}
where $h=u_a e B a/2$ and $\epsilon(k)=\sqrt{(u_a k)^2 + t_\perp^2}$. 
Taking the trace and doing the Matsubara sums leaves us with 
\begin{eqnarray}
\frac{\langle P_H^{(2)} \rangle} {L \langle j_c \rangle} &=&  - \frac{\pi \gamma B a } {4 u_c K_c u_a^2} \int_{-\infty}^{+\infty}  dk \frac{d}{dk}\left([n_F(\epsilon(k)-h) \right. \nonumber  \\ && \left. -n_F(-\epsilon(k)-h)] \frac{d\epsilon}{dk} \right) \nonumber \\ 
&=& \frac{\pi \gamma B a}{2 u_c K_c u_a}, 
\end{eqnarray}
independent of temperature, whereas 
\begin{eqnarray}
&&\frac{\langle P_H^{(1)} \rangle} {L \langle j_c \rangle} =  - \frac{\pi \gamma  } {2 e  u_c u_a K_c } \int_{-\infty}^{+\infty} dk k \frac{d}{dk} \left[n_F(\epsilon(k)-h)\right. \nonumber \\ && \left.  -n_F(-\epsilon(k)-h)\right] \\
&&= - \frac{\pi \gamma  } {2 e  u_c u_a K_c } \int_{-\infty}^{+\infty} dk [n_F(\epsilon(k)-h)-n_F(-\epsilon(k)-h)], \nonumber
\end{eqnarray}
contains the whole temperature dependence of $\langle P_H \rangle$.
The resulting Hall imbalance is 
\begin{eqnarray}\label{eq:imbalance-thermal}
&&\frac{\langle P_H \rangle}{L \langle j_c \rangle} =\frac{\pi \gamma}{2 e u_c u_a K_c } \left\{e B a - \int_{-\infty}^{+\infty}  n_F(\epsilon(k)-h) dk \right. \nonumber \\ && \left. -\int_{-\infty}^{+\infty} n_F(-\epsilon(k)-h) dk \right\}.   
\end{eqnarray}
The resulting temperature dependence is represented on Fig.~\ref{fig:LE_imbalance}. 
In the Meissner phase, $|h|<t_\perp$,  and $\langle P_H^{(1)}\rangle=O(e^{-\beta(t_\perp-|h|)})$ at low temperature, giving a near plateau,  whereas in the Vortex phase, the correction is $O(T^2)$ giving a faster drop of the Hall imbalance. In both cases, the low temperature Hall imbalance goes asymptotically to the ground state value.  At high temperature, $k_B T \gg t_\perp$ , an asymptotic expansion (see App.~\ref{app:expansion}) of Eq.~(\ref{eq:imbalance-thermal}) yields
\begin{equation}
\frac{\langle P_H \rangle}{L \langle j_c \rangle} = \frac{B a \pi \gamma}{2 u_c u_a K_c T^2} \left[\frac{7\zeta(3)}{4\pi^2} t_\perp^2 +\frac 1 3 \left(\frac{u_a e B a}{2}\right)^2\right] +o(T^{-2}),      
\end{equation}
leading to a vanishing Hall imbalance when $T\to +\infty$. {Let us remark that bosonization applies well when $t_\perp \ll \hbar u/a$ where $u$ is the sound velocity and $a$ is the lattice spacing. Thus in this regime the temperature behavior shown in Fig.\ref{fig:LE_imbalance} should be visible.}

We can also calculate the Hall voltage, obtained by adding a potential difference $V_H$ between the
two chains $V_H$ such that $\langle P_h \rangle=0.$ It's expression at finite temperature is:
\begin{equation}
V_H=\frac{\gamma \langle j_\rho\rangle }{2 e u_\rho K_\rho}\lbrack Ba-\frac 1 2 \int_{-\infty}^{+\infty} dk [n_F(\epsilon(k)-h)-n_F(-\epsilon(k)-h)], 
\end{equation}
thus we find again a suppression $1/T^2$ as a function of the temperature.

{Away from the Luther-Emery point, the quantum sine-Gordon model remains integrable\cite{essler04_condmat_exact_review}. Its spectrum consists of solitons, antisolitons and their bound states\cite{essler04_condmat_exact_review}. An exact calculation  of the Hall imbalance at finite temperature requires both   the correlator of the momentum of the sine-Gordon model with the gradient of the dual field  and the autocorrelator of the dual field gradient  in the presence of  chemical potentials for solitons and antisolitons. In principle, such a calculation could be attempted using the form factor expansion\cite{karowski_ff} generalized to finite temperature\cite{essler-konik_2009}, but would be very challenging technically.  Nevertheless, the Luther-Emery limit allows us already to gain a qualitative insight. In the ground state, with the soliton/antisoliton chemical potential below the gap, the autocorrelator of the dual field gradient is finite, but the momentum-dual field gradient correlator is zero.  At finite temperature, the  latter correlator can become finite, but because of the gap it will be suppressed by the Boltzmann factor. So even away from the Luther-Emery point, changes to the zero temperature Hall resistance will be exponentially suppressed at low temperatures. }
At high temperature, much larger than the ground state gap, a perturbative expansion can be applied to (\ref{eq:imbalance-prl}). At order zero, the correlator $\langle T_\tau \Pi_a \partial_x \phi_a (x,\tau) \phi_a(0,0)\rangle$ vanishes due to the symmetry $\phi_a \to -\phi_a$ and $\theta_a \to -\theta_a$. At low field, it is enough to expand to first order in $A_a$ in (\ref{eq:imbalance-prl}). This yields 
\begin{widetext}
\begin{eqnarray}
\frac{\langle P_H\rangle}{L \langle j_c \rangle} &=& \frac{\gamma \sqrt{2}}{e u_c K_c} \left[u_a K_a e A_a \int_0^\beta d\tau_1 \int dx_1 \int_0^\beta d\tau_2 \int dx_2 \langle T_\tau \pi [\Pi_a \partial_x \phi_a] (x_1,\tau_1) \Pi_a(x_2,\tau_2) \partial_x \phi_a(0,0) \rangle \right. \nonumber \\ 
&& \left. - e A_a \int_0^\beta \int dx \langle T_\tau \partial_x \phi_a(x,\tau) \partial_x \phi_a(0,0) \rangle\right]  + O(A_a)^2.  
\end{eqnarray}
\end{widetext} 
For $t_\perp=0$, the chains are decoupled, and the above expression must vanish. To find a non-trivial result, we need to expand in powers of $t_\perp$. The first order term is zero, and the lowest order result for the Hall imbalance is then 
\begin{eqnarray}
  \frac{\langle P_H\rangle}{L \langle j_c \rangle} &=& \frac{\gamma \sqrt{2} u_a K_a e A_a }{e u_c K_c} \left(\frac{2t_\perp}{a_0}\right)^2  \prod_{j=1}^4 \int_0^\beta d\tau_j \int dx_j    \nonumber \\ 
 && \times  \langle T_\tau [\pi \Pi_a \partial_x \phi_a] (x_1,\tau_1)  \Pi_a(x_2,\tau_2)  \cos \sqrt{2} \theta_a(x_3,\tau_3) \nonumber \\ 
 &&  \times \cos \sqrt{2} \theta_a(x_4,\tau_4) \partial_x \phi_a(0,0) \rangle. 
\end{eqnarray}
The above integral can now be estimated by a scaling argument. Each $\tau$ integration contributes a factor $\beta$, each $x$ integration a factor $u\beta$. Meanwhile, since $\Pi_a$ and $\partial_x \phi_a$ have scaling dimension $1$ and $\cos \sqrt{2} \theta_a$ has scaling dimension $1/(2K_a)$ the integrand contributes a factor $(u\beta/\alpha)^{-4-1/K_a}$. This leads to the estimate 
\begin{equation}
  \frac{\langle P_H\rangle}{L B \langle j_c \rangle} \sim T^{1/K_a-4}.    
\end{equation}
For $K_a=1/2$, the result at the Luther-Emery point is recovered. Away from that point, we expect the Hall imbalance to vanish at high temperature for $K_a >1/4$. 
\subsection{Three-legs ladder}
\label{ssec:three-legs}
With more than two chains, finding a Luther-Emery point becomes increasingly difficult. However, with 3 chains, this is still straightforward if we allow the Tomonaga-Luttinger exponent of $\phi_a$ and $\phi_b$ to differ and take the values $K_a=1/4$ and $K_b=3/4$, while the velocities $u_a=u_b$. Then, we can rescale the fields using $\left(\bar{\theta}_a=\theta_a/2;\bar{\phi}_a=2 \phi_a\right)$ and 
$\left(\bar{\theta}_b=\sqrt{\frac{3}{4}}\theta_b;\bar{\phi}_b=\sqrt{\frac{4}{3}}\phi_b\right)$  yielding in Eq.~(\ref{eq:hopping}) a term $2t_\perp \cos \sqrt{2} \bar{\theta}_a \cos \sqrt{2} \bar \theta_b$. Thus, we obtain\cite{schulz_losalamos} the equivalent of a Hamiltonian describing fermions with spin in a commensurate potential. At this Luther-Emery point, we can repeat the treatment above, leading to the same conclusion as in the two chain case, namely a Hall polarization $\langle P_H \rangle(T) \simeq\langle P_H \rangle(T=0)+O(e^{-\Delta/T})$ in the Meissner phase and in the limit of small temperature, and $\langle P_H \rangle \simeq\langle P_H \rangle(T=0)+O(T^2)$ in the Vortex phase, while $\langle P_H \rangle \sim T^{-2}$ at high temperature. It seems reasonable that a similar behavior is obtained with $N>3$ coupled chains.  

\section{Conclusion}
We have analyzed an N-leg bosonic ladder in the ground state and in the presence of flux, and found that at low flux, in the Meissner phase, the Hall resistance was proportional to the derivative of the logarithm of the Kohn stiffness with respect to the particle density, generalizing the result obtained in the two-leg case \cite{citro_hall_2025}. In the Vortex phase, the Hall voltage stops being proportional to the applied flux, so it can be used as a marker of the transition. We have also considered the finite temperature case for the two-leg ladder in the exactly solvable Luther-Emery limit, and we found that the deviations from the zero temperature results were suppressed exponentially in the Meissner phase, while they were only suppressed as power laws on the Vortex phase. A similar result holds also for the three-leg ladder with appropriate transformation. This shows that the result of \cite{citro_hall_2025} is robust, both with respect to temperature and number of chains. Those predictions can be compared with numerical simulations \cite{buser_hall_ladder} and experiments on  ultracold atoms \cite{guan2020}. Finally, it is worthwhile considering also the generalization of the results for two-leg fermionic ladder  in  \cite{citro_hall_2025} to the case of many legs. An interesting issue is the limit of the number of coupled chains going to infinity, realizing an anisotropic two dimensional array. If our formula remained applicable in that limit, it would predict a divergent Hall resistance changing sign across the Mott transition in the two-dimensional case. In \cite{citro_hall_2025}, we had predicted that the Hall resistance of a two-leg fermionic ladder would also behave as a derivative of the logarithm of the Kohn stiffness with respect to the particle density. Obviously, it would be interesting to see if, as in the bosonic case, this result is also valid with any number of coupled chains. Another open question is the role of interchain interactions. When those interactions dominate, they impose at zero flux a density wave ordering that competes with the Meissner phase \cite{lecheminant_exotic_2012}. Such transition could have a signature in the behavior of the Hall resistance. So far, we have only discussed cases where the charge mode could be treated as a perfect Tomonaga-Luttinger liquid, with a nonzero charge stiffness even at nonzero temperatures. However, in nonintegrable models, the charge stiffness vanishes as soon as the temperature is nonzero \cite{bertini-review_2021}, and the Drude peak in the conductivity is replaced by a Lorentzian. It would be worthwhile to investigate whether a relation between the weight of the Lorentzian and the Hall resistance can be established.  Models containing coupling to a bath \cite{caldeira_leggett,ambegaokar_josephson_dissipation_short,majumdar_bath2023} in the action or obeying a Lindbladian dynamics \cite{guo-lindblad_2017} could help clarifying that issue. 
Finally, let us stress that our results provide a bosonic multileg ladder perspective on Hall responses, establishing a  bridge to the recent experimental observations\cite{zhou_2022,zhou_2025}.
\begin{acknowledgments}
This research was supported in part by the Swiss National Science Foundation under grant 200020-219400.   E. O. and R. C. acknowledge support and hospitality from the university of Geneva. 
\end{acknowledgments}

\appendix
\section{Hahn polynomial basis for two or three chains}\label{app:hahn-lown}
Using the Rodrigues formula Eq.~(18.20.1) in \cite{olver_nist_2010}, we have
\begin{eqnarray}\label{eq:rodrigues}
 Q_n(X;0,0,N)=\frac{(-1)^n (N-n)!}{N! n!} \nabla_X^n \left[\prod_{l=0}^{n-1} (X+l+1)(X+l-N)\right] \nonumber \\   
\end{eqnarray}
where $\nabla_X F(X)=F(X)-F(X-1)$ is the backward difference operator. 
With two chains, we need the Hahn polynomials $Q_0(n,0,0,1)$ and $Q_1(n,0,0,1)$ with $n=0,1$. Using Eq.~(\ref{eq:rodrigues}), $Q_0(X,0,0,1)=1$ and $Q_1(X,0,0,1)=1-2X$. After normalization, 
\begin{equation}
    \left(\begin{array}{cc}
    q_0(1) & q_1(1) \\
    q_0(2) & q_1(2)  \\
    \end{array}\right) = \left(\begin{array}{cc}
    \frac 1 {\sqrt{2}} & \frac 1 {\sqrt{2}}  \\  
    \frac 1 {\sqrt{2}} & -\frac 1 {\sqrt{2}}  \\ 
    \end{array}\right),  
\end{equation}
and the symmetric/antisymmetric basis used in \cite{citro_hall_2025} is recovered with $\varphi_0=\phi_c$ and $\varphi_1=\phi_a$. 
With three chains, the required polynomials are $Q_{0,1,2}(n,0,0,2)$ with $n=0,1,2$. Applying Eq.~(\ref{eq:rodrigues}) we find $Q_0(X,0,0,2)=1$, $Q_1(X,0,0,2)=1-X$ and $Q_2(X,0,0,2)=1-6X+3X^2$. After calculating their values in $X=0,1,2$ and normalizing, we have 
\begin{equation}
    \left(\begin{array}{ccc}
    q_0(1) & q_1(1) & q_2(1) \\
    q_0(2) & q_1(2) & q_2(2) \\
    q_0(3) & q_1(3) & q_2(3) \\
    \end{array}\right) = \left(\begin{array}{ccc}
    \frac 1 {\sqrt{3}} & \frac 1 {\sqrt{2}} & \frac 1 {\sqrt{6}} \\ 
    \frac 1 {\sqrt{3}} & 0 & - \frac 2 {\sqrt{6}} \\ 
    \frac 1 {\sqrt{3}} & -\frac 1 {\sqrt{2}} & \frac 1 {\sqrt{6}} \\ 
    
    \end{array}\right),  
\end{equation}
and we recover the basis used in Sec.~\ref{sec:three-leg}, Eq.~\ref{eq:basis}, with $\varphi_0=\phi_c$, $\varphi_1=\phi_a$ and $\varphi_2=\phi_b$. 

\section{High temperature asymptotic expansion of the Hall imbalance}\label{app:expansion}
To derive the asymptotic expansion of the Hall imbalance, we start by rewriting 
\begin{eqnarray}
I(T)&=&\frac{B a} 2 -\frac 1 {2e} \int_{-\infty}^{+\infty} dk [n_F(\epsilon(k)-h)-n_F(-\epsilon(k)-h)], \nonumber \\
&=& \frac{Ba} 2 \left[1-\frac{2 \sinh\left(\frac{\beta u_a e B a}{2}\right) }{\beta u_a e B a} \right. \nonumber \\ 
&& \times\left.  \int_0^{+\infty} \frac{dy}{\cosh \sqrt{y^2 +(\beta t_\perp)^2} + \cosh\left(\frac{\beta u_a e B a}{2}\right)} \right]
\end{eqnarray}
In the limit $\beta \to 0$, the expression becomes 
\begin{equation}
    I(T)=\frac{Ba}2 \left(1-\int_0^{+\infty} \frac{dy}{1+\cosh y} \right) = 0.  
\end{equation}
This allows us to rewrite 
\begin{widetext}
\begin{eqnarray}
I(T)&=&\frac{Ba} 2 \int_0^{+\infty} \frac{dy}{(1+\cosh y) [\cosh
        \sqrt{y^2 +(\beta t_\perp)^2} + \cosh({\beta u_a e B a}/{2})]}
        \times \nonumber \\
    && \times \left[\cosh \sqrt{y^2 +(\beta t_\perp)^2} + \cosh\left(\frac{\beta u_a e B a}{2}\right)  - \frac{2 \sinh\left(\frac{\beta u_a e B a}{2}\right) }{\beta u_a e B a}     \times (1+\cosh y) \right]    
\end{eqnarray}
Using a Taylor expansion of $\cosh \sqrt{y^2 +(\beta t_\perp)^2}$ in powers of $\beta$, we obtain the approximation 
\begin{eqnarray}
 I(T)&=&\frac{Ba} 2 \int_0^{+\infty} \frac{dy}{(1+\cosh y)^2} \left[\left(\frac 2 3 + \frac 1 6 \cosh y \right)\left(\frac{\beta u_a e B a}{2}\right)^2 + \frac{(\beta t_\perp)}2 \frac {\sinh y}{y} \right] + O(\beta^{4})\nonumber \\
  &=& \frac{Ba} 2 \left[ \frac 1 3 \left(\frac{\beta u_a e B a}{2}\right)^2  + \frac{7\zeta(3)}{4\pi^2} (\beta t_\perp)^2  \right] + O(\beta^{4})
\end{eqnarray}
\end{widetext}


\begin{thebibliography}{93}%
\makeatletter
\providecommand \@ifxundefined [1]{%
 \@ifx{#1\undefined}
}%
\providecommand \@ifnum [1]{%
 \ifnum #1\expandafter \@firstoftwo
 \else \expandafter \@secondoftwo
 \fi
}%
\providecommand \@ifx [1]{%
 \ifx #1\expandafter \@firstoftwo
 \else \expandafter \@secondoftwo
 \fi
}%
\providecommand \natexlab [1]{#1}%
\providecommand \enquote  [1]{``#1''}%
\providecommand \bibnamefont  [1]{#1}%
\providecommand \bibfnamefont [1]{#1}%
\providecommand \citenamefont [1]{#1}%
\providecommand \href@noop [0]{\@secondoftwo}%
\providecommand \href [0]{\begingroup \@sanitize@url \@href}%
\providecommand \@href[1]{\@@startlink{#1}\@@href}%
\providecommand \@@href[1]{\endgroup#1\@@endlink}%
\providecommand \@sanitize@url [0]{\catcode `\\12\catcode `\$12\catcode
  `\&12\catcode `\#12\catcode `\^12\catcode `\_12\catcode `\%12\relax}%
\providecommand \@@startlink[1]{}%
\providecommand \@@endlink[0]{}%
\providecommand \url  [0]{\begingroup\@sanitize@url \@url }%
\providecommand \@url [1]{\endgroup\@href {#1}{\urlprefix }}%
\providecommand \urlprefix  [0]{URL }%
\providecommand \Eprint [0]{\href }%
\providecommand \doibase [0]{https://doi.org/}%
\providecommand \selectlanguage [0]{\@gobble}%
\providecommand \bibinfo  [0]{\@secondoftwo}%
\providecommand \bibfield  [0]{\@secondoftwo}%
\providecommand \translation [1]{[#1]}%
\providecommand \BibitemOpen [0]{}%
\providecommand \bibitemStop [0]{}%
\providecommand \bibitemNoStop [0]{.\EOS\space}%
\providecommand \EOS [0]{\spacefactor3000\relax}%
\providecommand \BibitemShut  [1]{\csname bibitem#1\endcsname}%
\let\auto@bib@innerbib\@empty
\bibitem [{\citenamefont {Hall}(1879)}]{Hall1879}%
  \BibitemOpen
  \bibfield  {author} {\bibinfo {author} {\bibfnamefont {E.~H.}\ \bibnamefont
  {Hall}},\ }\bibfield  {title} {\bibinfo {title} {On a new action of the
  magnet on electric currents},\ }\href
  {http://www.ajsonline.org/content/s3-19/111/200.short} {\bibfield  {journal}
  {\bibinfo  {journal} {Am. J. Math.}\ }\textbf {\bibinfo {volume} {2}},\
  \bibinfo {pages} {287} (\bibinfo {year} {1879})}\BibitemShut {NoStop}%
\bibitem [{\citenamefont {von Klitzing}\ \emph {et~al.}(1980)\citenamefont {von
  Klitzing}, \citenamefont {Dorda},\ and\ \citenamefont
  {Pepper}}]{klitzing_IQHE}%
  \BibitemOpen
  \bibfield  {author} {\bibinfo {author} {\bibfnamefont {K.}~\bibnamefont {von
  Klitzing}}, \bibinfo {author} {\bibfnamefont {G.}~\bibnamefont {Dorda}},\
  and\ \bibinfo {author} {\bibfnamefont {M.}~\bibnamefont {Pepper}},\
  }\bibfield  {title} {\bibinfo {title} {{No Title}},\ }\href@noop {}
  {\bibfield  {journal} {\bibinfo  {journal} {Physical Review Letters}\
  }\textbf {\bibinfo {volume} {45}},\ \bibinfo {pages} {494} (\bibinfo {year}
  {1980})}\BibitemShut {NoStop}%
\bibitem [{\citenamefont {Klitzing}\ \emph {et~al.}(0 08)\citenamefont
  {Klitzing}, \citenamefont {Dorda},\ and\ \citenamefont
  {Pepper}}]{Klitzing1980}%
  \BibitemOpen
  \bibfield  {author} {\bibinfo {author} {\bibfnamefont {K.~v.}\ \bibnamefont
  {Klitzing}}, \bibinfo {author} {\bibfnamefont {G.}~\bibnamefont {Dorda}},\
  and\ \bibinfo {author} {\bibfnamefont {M.}~\bibnamefont {Pepper}},\
  }\bibfield  {title} {\bibinfo {title} {New method for high-accuracy
  determination of the fine-structure constant based on quantized hall
  resistance},\ }\href {https://doi.org/10.1103/PhysRevLett.45.494} {\bibfield
  {journal} {\bibinfo  {journal} {Phys. Rev. Lett.}\ }\textbf {\bibinfo
  {volume} {45}},\ \bibinfo {pages} {494} (\bibinfo {year}
  {1980-08})}\BibitemShut {NoStop}%
\bibitem [{\citenamefont {Ripka}\ and\ \citenamefont
  {Z{\`a}v{\v{e}}ta}(2009)}]{ripka2009}%
  \BibitemOpen
  \bibfield  {author} {\bibinfo {author} {\bibfnamefont {P.}~\bibnamefont
  {Ripka}}\ and\ \bibinfo {author} {\bibfnamefont {K.}~\bibnamefont
  {Z{\`a}v{\v{e}}ta}},\ }\bibfield  {title} {\bibinfo {title} {Magnetic
  sensors: Principles and applications},\ }in\ \href
  {https://doi.org/https://doi.org/10.1016/S1567-2719(09)01803-4} {\emph
  {\bibinfo {booktitle} {Handbook of Magnetic Materials}}},\ \bibinfo {series}
  {Handbook of Magnetic Materials}, Vol.~\bibinfo {volume} {18},\ \bibinfo
  {editor} {edited by\ \bibinfo {editor} {\bibfnamefont {K.}~\bibnamefont
  {Buschow}}}\ (\bibinfo  {publisher} {Elsevier},\ \bibinfo {year} {2009})\
  Chap.~\bibinfo {chapter} {3}, p.\ \bibinfo {pages} {347}\BibitemShut
  {NoStop}%
\bibitem [{\citenamefont {Thouless}\ \emph {et~al.}(1982)\citenamefont
  {Thouless}, \citenamefont {Kohmoto}, \citenamefont {Nightingale},\ and\
  \citenamefont {den Nijs}}]{thouless_1982}%
  \BibitemOpen
  \bibfield  {author} {\bibinfo {author} {\bibfnamefont {D.~J.}\ \bibnamefont
  {Thouless}}, \bibinfo {author} {\bibfnamefont {M.}~\bibnamefont {Kohmoto}},
  \bibinfo {author} {\bibfnamefont {M.~P.}\ \bibnamefont {Nightingale}},\ and\
  \bibinfo {author} {\bibfnamefont {M.}~\bibnamefont {den Nijs}},\ }\bibfield
  {title} {\bibinfo {title} {Quantized hall conductance in a two-dimensional
  periodic potential},\ }\href {https://doi.org/10.1103/PhysRevLett.49.405}
  {\bibfield  {journal} {\bibinfo  {journal} {Phys. Rev. Lett.}\ }\textbf
  {\bibinfo {volume} {49}},\ \bibinfo {pages} {405} (\bibinfo {year}
  {1982})}\BibitemShut {NoStop}%
\bibitem [{\citenamefont {Avron}\ and\ \citenamefont {Seiler}(5
  01)}]{Avron1985}%
  \BibitemOpen
  \bibfield  {author} {\bibinfo {author} {\bibfnamefont {J.~E.}\ \bibnamefont
  {Avron}}\ and\ \bibinfo {author} {\bibfnamefont {R.}~\bibnamefont {Seiler}},\
  }\bibfield  {title} {\bibinfo {title} {Quantization of the hall conductance
  for general, multiparticle schr\"odinger hamiltonians},\ }\href
  {https://doi.org/10.1103/PhysRevLett.54.259} {\bibfield  {journal} {\bibinfo
  {journal} {Phys. Rev. Lett.}\ }\textbf {\bibinfo {volume} {54}},\ \bibinfo
  {pages} {259} (\bibinfo {year} {1985-01})}\BibitemShut {NoStop}%
\bibitem [{\citenamefont {Niu}\ \emph {et~al.}(5 03)\citenamefont {Niu},
  \citenamefont {Thouless},\ and\ \citenamefont {Wu}}]{Niu1985}%
  \BibitemOpen
  \bibfield  {author} {\bibinfo {author} {\bibfnamefont {Q.}~\bibnamefont
  {Niu}}, \bibinfo {author} {\bibfnamefont {D.~J.}\ \bibnamefont {Thouless}},\
  and\ \bibinfo {author} {\bibfnamefont {Y.-S.}\ \bibnamefont {Wu}},\
  }\bibfield  {title} {\bibinfo {title} {Quantized hall conductance as a
  topological invariant},\ }\href {https://doi.org/10.1103/PhysRevB.31.3372}
  {\bibfield  {journal} {\bibinfo  {journal} {Phys. Rev. B}\ }\textbf {\bibinfo
  {volume} {31}},\ \bibinfo {pages} {3372} (\bibinfo {year}
  {1985-03})}\BibitemShut {NoStop}%
\bibitem [{\citenamefont {Xiao}\ \emph {et~al.}(2010)\citenamefont {Xiao},
  \citenamefont {Chang},\ and\ \citenamefont {Niu}}]{review_niu_2010}%
  \BibitemOpen
  \bibfield  {author} {\bibinfo {author} {\bibfnamefont {D.}~\bibnamefont
  {Xiao}}, \bibinfo {author} {\bibfnamefont {M.-C.}\ \bibnamefont {Chang}},\
  and\ \bibinfo {author} {\bibfnamefont {Q.}~\bibnamefont {Niu}},\ }\bibfield
  {title} {\bibinfo {title} {Berry phase effects on electronic properties},\
  }\href {https://doi.org/10.1103/RevModPhys.82.1959} {\bibfield  {journal}
  {\bibinfo  {journal} {Rev. Mod. Phys.}\ }\textbf {\bibinfo {volume} {82}},\
  \bibinfo {pages} {1959} (\bibinfo {year} {2010})}\BibitemShut {NoStop}%
\bibitem [{\citenamefont {Tsuji}(1958)}]{Tsuji1958a}%
  \BibitemOpen
  \bibfield  {author} {\bibinfo {author} {\bibfnamefont {M.}~\bibnamefont
  {Tsuji}},\ }\bibfield  {title} {\bibinfo {title} {The thermoelectric,
  galvanomagnetic and thermomagnetic effects of monovalent metals. iii. the
  galvanomagnetic and thermomagnetic effects for anisotropic media},\
  }\href@noop {} {\bibfield  {journal} {\bibinfo  {journal} {Journal of the
  Physical Society of Japan}\ }\textbf {\bibinfo {volume} {13}},\ \bibinfo
  {pages} {979} (\bibinfo {year} {1958})}\BibitemShut {NoStop}%
\bibitem [{\citenamefont {Ong}(1991)}]{Ong1991a}%
  \BibitemOpen
  \bibfield  {author} {\bibinfo {author} {\bibfnamefont {N.}~\bibnamefont
  {Ong}},\ }\bibfield  {title} {\bibinfo {title} {Geometric interpretation of
  the weak-field hall conductivity in two-dimensional metals with arbitrary
  fermi surface},\ }\href@noop {} {\bibfield  {journal} {\bibinfo  {journal}
  {Physical Review B}\ }\textbf {\bibinfo {volume} {43}},\ \bibinfo {pages}
  {193} (\bibinfo {year} {1991})}\BibitemShut {NoStop}%
\bibitem [{\citenamefont {Le\'{o}n}\ \emph {et~al.}(2007)\citenamefont
  {Le\'{o}n}, \citenamefont {Berthod},\ and\ \citenamefont
  {Giamarchi}}]{Leon-PRB-2007}%
  \BibitemOpen
  \bibfield  {author} {\bibinfo {author} {\bibfnamefont {G.}~\bibnamefont
  {Le\'{o}n}}, \bibinfo {author} {\bibfnamefont {C.}~\bibnamefont {Berthod}},\
  and\ \bibinfo {author} {\bibfnamefont {T.}~\bibnamefont {Giamarchi}},\
  }\bibfield  {title} {\bibinfo {title} {Hall effect in strongly correlated
  low-dimensional systems},\ }\href@noop {} {\bibfield  {journal} {\bibinfo
  {journal} {Physical Review B}\ }\textbf {\bibinfo {volume} {75}},\ \bibinfo
  {pages} {195123} (\bibinfo {year} {2007})}\BibitemShut {NoStop}%
\bibitem [{\citenamefont {Zotos}\ \emph {et~al.}(2000)\citenamefont {Zotos},
  \citenamefont {Naef}, \citenamefont {Long},\ and\ \citenamefont
  {Prelov\ifmmode~\check{s}\else \v{s}\fi{}ek}}]{zotos_2000}%
  \BibitemOpen
  \bibfield  {author} {\bibinfo {author} {\bibfnamefont {X.}~\bibnamefont
  {Zotos}}, \bibinfo {author} {\bibfnamefont {F.}~\bibnamefont {Naef}},
  \bibinfo {author} {\bibfnamefont {M.}~\bibnamefont {Long}},\ and\ \bibinfo
  {author} {\bibfnamefont {P.}~\bibnamefont {Prelov\ifmmode~\check{s}\else
  \v{s}\fi{}ek}},\ }\bibfield  {title} {\bibinfo {title} {Reactive hall
  response},\ }\href {https://doi.org/10.1103/PhysRevLett.85.377} {\bibfield
  {journal} {\bibinfo  {journal} {Phys. Rev. Lett.}\ }\textbf {\bibinfo
  {volume} {85}},\ \bibinfo {pages} {377} (\bibinfo {year} {2000})}\BibitemShut
  {NoStop}%
\bibitem [{\citenamefont {Auerbach}(2018)}]{Auerbach2018}%
  \BibitemOpen
  \bibfield  {author} {\bibinfo {author} {\bibfnamefont {A.}~\bibnamefont
  {Auerbach}},\ }\bibfield  {title} {\bibinfo {title} {Hall number of strongly
  correlated metals},\ }\href {https://doi.org/10.1103/PhysRevLett.121.066601}
  {\bibfield  {journal} {\bibinfo  {journal} {Phys. Rev. Lett.}\ }\textbf
  {\bibinfo {volume} {121}},\ \bibinfo {pages} {066601} (\bibinfo {year}
  {2018})}\BibitemShut {NoStop}%
\bibitem [{\citenamefont {Shastry}\ \emph
  {et~al.}(1993{\natexlab{a}})\citenamefont {Shastry}, \citenamefont
  {Shraiman},\ and\ \citenamefont {Singh}}]{shastry_Hall}%
  \BibitemOpen
  \bibfield  {author} {\bibinfo {author} {\bibfnamefont {B.~S.}\ \bibnamefont
  {Shastry}}, \bibinfo {author} {\bibfnamefont {B.~I.}\ \bibnamefont
  {Shraiman}},\ and\ \bibinfo {author} {\bibfnamefont {R.~R.}\ \bibnamefont
  {Singh}},\ }\href@noop {} {\bibfield  {journal} {\bibinfo  {journal}
  {Physical Review Letters}\ }\textbf {\bibinfo {volume} {70}},\ \bibinfo
  {pages} {2004} (\bibinfo {year} {1993}{\natexlab{a}})}\BibitemShut {NoStop}%
\bibitem [{\citenamefont {Fazio}\ and\ \citenamefont {van~der
  Zant}(2001)}]{fazio_josephson_junction_review}%
  \BibitemOpen
  \bibfield  {author} {\bibinfo {author} {\bibfnamefont {R.}~\bibnamefont
  {Fazio}}\ and\ \bibinfo {author} {\bibfnamefont {H.}~\bibnamefont {van~der
  Zant}},\ }\href@noop {} {\bibfield  {journal} {\bibinfo  {journal} {Physics
  Reports}\ }\textbf {\bibinfo {volume} {355}},\ \bibinfo {pages} {235}
  (\bibinfo {year} {2001})}\BibitemShut {NoStop}%
\bibitem [{\citenamefont {Dalibard}\ \emph {et~al.}(2011)\citenamefont
  {Dalibard}, \citenamefont {Gerbier}, \citenamefont
  {Juzeli\ifmmode~\bar{u}\else \={u}\fi{}nas},\ and\ \citenamefont
  {\"Ohberg}}]{dalibard_review_artificial_gauge_fields}%
  \BibitemOpen
  \bibfield  {author} {\bibinfo {author} {\bibfnamefont {J.}~\bibnamefont
  {Dalibard}}, \bibinfo {author} {\bibfnamefont {F.}~\bibnamefont {Gerbier}},
  \bibinfo {author} {\bibfnamefont {G.}~\bibnamefont
  {Juzeli\ifmmode~\bar{u}\else \={u}\fi{}nas}},\ and\ \bibinfo {author}
  {\bibfnamefont {P.}~\bibnamefont {\"Ohberg}},\ }\bibfield  {title} {\bibinfo
  {title} {\textit{Colloquium} : Artificial gauge potentials for neutral
  atoms},\ }\href {https://doi.org/10.1103/RevModPhys.83.1523} {\bibfield
  {journal} {\bibinfo  {journal} {Rev. Mod. Phys.}\ }\textbf {\bibinfo {volume}
  {83}},\ \bibinfo {pages} {1523} (\bibinfo {year} {2011})}\BibitemShut
  {NoStop}%
\bibitem [{\citenamefont {Galitski}\ and\ \citenamefont
  {Spielman}(2013)}]{Galitski2013}%
  \BibitemOpen
  \bibfield  {author} {\bibinfo {author} {\bibfnamefont {V.}~\bibnamefont
  {Galitski}}\ and\ \bibinfo {author} {\bibfnamefont {I.}~\bibnamefont
  {Spielman}},\ }\bibfield  {title} {\bibinfo {title} {Spin–orbit coupling in
  quantum gases},\ }\href@noop {} {\bibfield  {journal} {\bibinfo  {journal}
  {Nature (London)}\ }\textbf {\bibinfo {volume} {494}},\ \bibinfo {pages} {49}
  (\bibinfo {year} {2013})}\BibitemShut {NoStop}%
\bibitem [{\citenamefont {Mancini}\ \emph {et~al.}(2015)\citenamefont
  {Mancini}, \citenamefont {Pagano}, \citenamefont {Cappellini}, \citenamefont
  {Livi}, \citenamefont {Rider}, \citenamefont {Catani}, \citenamefont {Sias},
  \citenamefont {Zoller}, \citenamefont {Inguscio}, \citenamefont {Dalmonte},\
  and\ \citenamefont {Fallani}}]{mancini_observation_2015}%
  \BibitemOpen
  \bibfield  {author} {\bibinfo {author} {\bibfnamefont {M.}~\bibnamefont
  {Mancini}}, \bibinfo {author} {\bibfnamefont {G.}~\bibnamefont {Pagano}},
  \bibinfo {author} {\bibfnamefont {G.}~\bibnamefont {Cappellini}}, \bibinfo
  {author} {\bibfnamefont {L.}~\bibnamefont {Livi}}, \bibinfo {author}
  {\bibfnamefont {M.}~\bibnamefont {Rider}}, \bibinfo {author} {\bibfnamefont
  {J.}~\bibnamefont {Catani}}, \bibinfo {author} {\bibfnamefont
  {C.}~\bibnamefont {Sias}}, \bibinfo {author} {\bibfnamefont {P.}~\bibnamefont
  {Zoller}}, \bibinfo {author} {\bibfnamefont {M.}~\bibnamefont {Inguscio}},
  \bibinfo {author} {\bibfnamefont {M.}~\bibnamefont {Dalmonte}},\ and\
  \bibinfo {author} {\bibfnamefont {L.}~\bibnamefont {Fallani}},\ }\bibfield
  {title} {\bibinfo {title} {Observation of chiral edge states with neutral
  fermions in synthetic {Hall} ribbons},\ }\href
  {https://doi.org/10.1126/science.aaa8736} {\bibfield  {journal} {\bibinfo
  {journal} {Science}\ }\textbf {\bibinfo {volume} {349}},\ \bibinfo {pages}
  {1510} (\bibinfo {year} {2015})}\BibitemShut {NoStop}%
\bibitem [{\citenamefont {Dalibard}(2016)}]{dalibard_gauge_2015}%
  \BibitemOpen
  \bibfield  {author} {\bibinfo {author} {\bibfnamefont {J.}~\bibnamefont
  {Dalibard}},\ }\bibfield  {title} {\bibinfo {title} {Introduction to the
  physics of artificial gauge fields},\ }in\ \href
  {http://arxiv.org/abs/1504.05520} {\emph {\bibinfo {booktitle} {Quantum
  {Matter} at {Ultralow} {Temperatures}}}},\ \bibinfo {editor} {edited by\
  \bibinfo {editor} {\bibfnamefont {M.}~\bibnamefont {Inguscio}}, \bibinfo
  {editor} {\bibfnamefont {W.}~\bibnamefont {Ketterle}}, \bibinfo {editor}
  {\bibfnamefont {S.}~\bibnamefont {Stringari}},\ and\ \bibinfo {editor}
  {\bibfnamefont {G.}~\bibnamefont {Roati}}}\ (\bibinfo  {publisher} {IOS
  Press},\ \bibinfo {address} {Amsterdam},\ \bibinfo {year} {2016})\
  p.~\bibinfo {pages} {1},\ \bibinfo {note} {arXiv:1504.05520
  [cond-mat]}\BibitemShut {NoStop}%
\bibitem [{\citenamefont {Goldman}\ \emph {et~al.}(2016)\citenamefont
  {Goldman}, \citenamefont {Budich},\ and\ \citenamefont
  {Zoller}}]{Goldman2016}%
  \BibitemOpen
  \bibfield  {author} {\bibinfo {author} {\bibfnamefont {N.}~\bibnamefont
  {Goldman}}, \bibinfo {author} {\bibfnamefont {J.}~\bibnamefont {Budich}},\
  and\ \bibinfo {author} {\bibfnamefont {P.}~\bibnamefont {Zoller}},\
  }\bibfield  {title} {\bibinfo {title} {Topological quantum matter with
  ultracold gases in optical lattices},\ }\href@noop {} {\bibfield  {journal}
  {\bibinfo  {journal} {Nat. Phys.}\ }\textbf {\bibinfo {volume} {12}},\
  \bibinfo {pages} {639} (\bibinfo {year} {2016})}\BibitemShut {NoStop}%
\bibitem [{\citenamefont {Stuhl}\ \emph {et~al.}(2015)\citenamefont {Stuhl},
  \citenamefont {Lu}, \citenamefont {Aycock}, \citenamefont {Genkina},\ and\
  \citenamefont {Spielman}}]{stuhl_ladder_2015}%
  \BibitemOpen
  \bibfield  {author} {\bibinfo {author} {\bibfnamefont {B.~K.}\ \bibnamefont
  {Stuhl}}, \bibinfo {author} {\bibfnamefont {H.-I.}\ \bibnamefont {Lu}},
  \bibinfo {author} {\bibfnamefont {L.~M.}\ \bibnamefont {Aycock}}, \bibinfo
  {author} {\bibfnamefont {D.}~\bibnamefont {Genkina}},\ and\ \bibinfo {author}
  {\bibfnamefont {I.~B.}\ \bibnamefont {Spielman}},\ }\bibfield  {title}
  {\bibinfo {title} {Visualizing edge states with an atomic bose gas in the
  quantum hall regime},\ }\href {https://doi.org/10.1126/science.aaa8515}
  {\bibfield  {journal} {\bibinfo  {journal} {Science}\ }\textbf {\bibinfo
  {volume} {349}},\ \bibinfo {pages} {1514} (\bibinfo {year}
  {2015})}\BibitemShut {NoStop}%
\bibitem [{\citenamefont {Zhou}\ \emph
  {et~al.}(2023{\natexlab{a}})\citenamefont {Zhou}, \citenamefont {Cappellini},
  \citenamefont {Tusi}, \citenamefont {Franchi}, \citenamefont {Beller},
  \citenamefont {Masini}, \citenamefont {Parravicini}, \citenamefont
  {Repellin}, \citenamefont {Greschner}, \citenamefont {Inguscio} \emph
  {et~al.}}]{zhou2023strongly}%
  \BibitemOpen
  \bibfield  {author} {\bibinfo {author} {\bibfnamefont {T.}~\bibnamefont
  {Zhou}}, \bibinfo {author} {\bibfnamefont {G.}~\bibnamefont {Cappellini}},
  \bibinfo {author} {\bibfnamefont {D.}~\bibnamefont {Tusi}}, \bibinfo {author}
  {\bibfnamefont {L.}~\bibnamefont {Franchi}}, \bibinfo {author} {\bibfnamefont
  {T.}~\bibnamefont {Beller}}, \bibinfo {author} {\bibfnamefont
  {G.}~\bibnamefont {Masini}}, \bibinfo {author} {\bibfnamefont
  {J.}~\bibnamefont {Parravicini}}, \bibinfo {author} {\bibfnamefont
  {C.}~\bibnamefont {Repellin}}, \bibinfo {author} {\bibfnamefont
  {S.}~\bibnamefont {Greschner}}, \bibinfo {author} {\bibfnamefont
  {M.}~\bibnamefont {Inguscio}}, \emph {et~al.},\ }\bibfield  {title} {\bibinfo
  {title} {Strongly interacting lattice fermions with coherent state
  manipulation: from universal hall response to hall voltage measurement},\
  }in\ \href@noop {} {\emph {\bibinfo {booktitle} {Bose-Einstein Condensation
  2023 Report of Contributions}}}\ (\bibinfo  {publisher} {Universit{\"a}t
  Hamburg},\ \bibinfo {year} {2023})\ pp.\ \bibinfo {pages} {8--8}\BibitemShut
  {NoStop}%
\bibitem [{\citenamefont {Impertro}\ \emph {et~al.}(2025)\citenamefont
  {Impertro}, \citenamefont {Huh}, \citenamefont {Karch}, \citenamefont
  {Wienand}, \citenamefont {Bloch},\ and\ \citenamefont
  {Aidelsburger}}]{impertro_strongly_2025}%
  \BibitemOpen
  \bibfield  {author} {\bibinfo {author} {\bibfnamefont {A.}~\bibnamefont
  {Impertro}}, \bibinfo {author} {\bibfnamefont {S.}~\bibnamefont {Huh}},
  \bibinfo {author} {\bibfnamefont {S.}~\bibnamefont {Karch}}, \bibinfo
  {author} {\bibfnamefont {J.~F.}\ \bibnamefont {Wienand}}, \bibinfo {author}
  {\bibfnamefont {I.}~\bibnamefont {Bloch}},\ and\ \bibinfo {author}
  {\bibfnamefont {M.}~\bibnamefont {Aidelsburger}},\ }\bibfield  {title}
  {\bibinfo {title} {Strongly interacting {Meissner} phases in large bosonic
  flux ladders},\ }\href {https://doi.org/10.1038/s41567-025-02890-0}
  {\bibfield  {journal} {\bibinfo  {journal} {Nature Physics}\ ,\ \bibinfo
  {pages} {1}} (\bibinfo {year} {2025})},\ \bibinfo {note} {arXiv:2412.09481
  [cond-mat]}\BibitemShut {NoStop}%
\bibitem [{\citenamefont {Haldane}(1981{\natexlab{a}})}]{haldane_bosons}%
  \BibitemOpen
  \bibfield  {author} {\bibinfo {author} {\bibfnamefont {F.~D.~M.}\
  \bibnamefont {Haldane}},\ }\href@noop {} {\bibfield  {journal} {\bibinfo
  {journal} {Physical Review Letters}\ }\textbf {\bibinfo {volume} {47}},\
  \bibinfo {pages} {1840} (\bibinfo {year} {1981}{\natexlab{a}})}\BibitemShut
  {NoStop}%
\bibitem [{\citenamefont {Efetov}\ and\ \citenamefont
  {Larkin}(1975)}]{efetov_larkin75}%
  \BibitemOpen
  \bibfield  {author} {\bibinfo {author} {\bibfnamefont {K.~B.}\ \bibnamefont
  {Efetov}}\ and\ \bibinfo {author} {\bibfnamefont {A.~I.}\ \bibnamefont
  {Larkin}},\ }\href@noop {} {\bibfield  {journal} {\bibinfo  {journal} {Sov.
  Phys. JETP}\ }\textbf {\bibinfo {volume} {42}},\ \bibinfo {pages} {390}
  (\bibinfo {year} {1975})}\BibitemShut {NoStop}%
\bibitem [{\citenamefont {Cazalilla}\ \emph {et~al.}(2011)\citenamefont
  {Cazalilla}, \citenamefont {Citro}, \citenamefont {Giamarchi}, \citenamefont
  {Orignac},\ and\ \citenamefont {Rigol}}]{cazalilla_review_bosons}%
  \BibitemOpen
  \bibfield  {author} {\bibinfo {author} {\bibfnamefont {M.~A.}\ \bibnamefont
  {Cazalilla}}, \bibinfo {author} {\bibfnamefont {R.}~\bibnamefont {Citro}},
  \bibinfo {author} {\bibfnamefont {T.}~\bibnamefont {Giamarchi}}, \bibinfo
  {author} {\bibfnamefont {E.}~\bibnamefont {Orignac}},\ and\ \bibinfo {author}
  {\bibfnamefont {M.}~\bibnamefont {Rigol}},\ }\bibfield  {title} {\bibinfo
  {title} {One dimensional bosons: From condensed matter systems to ultracold
  gases},\ }\href {https://doi.org/10.1103/RevModPhys.83.1405} {\bibfield
  {journal} {\bibinfo  {journal} {Rev. Mod. Phys.}\ }\textbf {\bibinfo {volume}
  {83}},\ \bibinfo {pages} {1405} (\bibinfo {year} {2011})}\BibitemShut
  {NoStop}%
\bibitem [{\citenamefont {White}(1992)}]{white_dmrg_letter}%
  \BibitemOpen
  \bibfield  {author} {\bibinfo {author} {\bibfnamefont {S.~R.}\ \bibnamefont
  {White}},\ }\href@noop {} {\bibfield  {journal} {\bibinfo  {journal}
  {Physical Review Letters}\ }\textbf {\bibinfo {volume} {69}},\ \bibinfo
  {pages} {2863} (\bibinfo {year} {1992})}\BibitemShut {NoStop}%
\bibitem [{\citenamefont {Hallberg}(2006)}]{Hallberg_rev}%
  \BibitemOpen
  \bibfield  {author} {\bibinfo {author} {\bibfnamefont {K.~A.}\ \bibnamefont
  {Hallberg}},\ }\href@noop {} {\bibfield  {journal} {\bibinfo  {journal} {adv.
  Phys.}\ }\textbf {\bibinfo {volume} {55}},\ \bibinfo {pages} {477} (\bibinfo
  {year} {2006})}\BibitemShut {NoStop}%
\bibitem [{\citenamefont {Schollw\"ock}(2011)}]{schollwock_DMRG_review}%
  \BibitemOpen
  \bibfield  {author} {\bibinfo {author} {\bibfnamefont {U.}~\bibnamefont
  {Schollw\"ock}},\ }\bibfield  {title} {\bibinfo {title} {The density-matrix
  renormalization group in the age of matrix product states},\ }\href@noop {}
  {\bibfield  {journal} {\bibinfo  {journal} {Annals of Physics}\ }\textbf
  {\bibinfo {volume} {326}},\ \bibinfo {pages} {96} (\bibinfo {year}
  {2011})}\BibitemShut {NoStop}%
\bibitem [{\citenamefont {{Kardar}}(1986)}]{kardar_josephson_ladder}%
  \BibitemOpen
  \bibfield  {author} {\bibinfo {author} {\bibfnamefont {M.}~\bibnamefont
  {{Kardar}}},\ }\bibfield  {title} {\bibinfo {title} {Josephson-junction
  ladders and quantum fluctuations},\ }\href
  {https://doi.org/10.1103/PhysRevB.33.3125} {\bibfield  {journal} {\bibinfo
  {journal} {Physical Review B}\ }\textbf {\bibinfo {volume} {33}},\ \bibinfo
  {pages} {3125} (\bibinfo {year} {1986})}\BibitemShut {NoStop}%
\bibitem [{\citenamefont {Orignac}\ and\ \citenamefont
  {Giamarchi}(2001)}]{orignac01_meissner}%
  \BibitemOpen
  \bibfield  {author} {\bibinfo {author} {\bibfnamefont {E.}~\bibnamefont
  {Orignac}}\ and\ \bibinfo {author} {\bibfnamefont {T.}~\bibnamefont
  {Giamarchi}},\ }\bibfield  {title} {\bibinfo {title} {Meissner effect in a
  bosonic ladder},\ }\href@noop {} {\bibfield  {journal} {\bibinfo  {journal}
  {Physical Review B}\ }\textbf {\bibinfo {volume} {64}},\ \bibinfo {pages}
  {144515} (\bibinfo {year} {2001})},\ \Eprint
  {https://arxiv.org/abs/cond-mat/0011497} {cond-mat/0011497} \BibitemShut
  {NoStop}%
\bibitem [{\citenamefont {Tokuno}\ and\ \citenamefont
  {Georges}(2014)}]{Tokuno2014}%
  \BibitemOpen
  \bibfield  {author} {\bibinfo {author} {\bibfnamefont {A.}~\bibnamefont
  {Tokuno}}\ and\ \bibinfo {author} {\bibfnamefont {A.}~\bibnamefont
  {Georges}},\ }\bibfield  {title} {\bibinfo {title} {Ground states of a
  bose-hubbard ladder in an artificial magnetic field: field-theoretical
  approach},\ }\href@noop {} {\bibfield  {journal} {\bibinfo  {journal} {New J.
  Phys.}\ }\textbf {\bibinfo {volume} {16}},\ \bibinfo {pages} {073005}
  (\bibinfo {year} {2014})}\BibitemShut {NoStop}%
\bibitem [{\citenamefont {Uchino}(2016)}]{Uchino2016a}%
  \BibitemOpen
  \bibfield  {author} {\bibinfo {author} {\bibfnamefont {S.}~\bibnamefont
  {Uchino}},\ }\bibfield  {title} {\bibinfo {title} {Analytical approach to a
  bosonic ladder subject to a magnetic field},\ }\href@noop {} {\bibfield
  {journal} {\bibinfo  {journal} {Phys. Rev. A}\ }\textbf {\bibinfo {volume}
  {93}},\ \bibinfo {pages} {053629} (\bibinfo {year} {2016})}\BibitemShut
  {NoStop}%
\bibitem [{\citenamefont {Cha}\ and\ \citenamefont {Shin}(2011)}]{cha2011}%
  \BibitemOpen
  \bibfield  {author} {\bibinfo {author} {\bibfnamefont {M.-C.}\ \bibnamefont
  {Cha}}\ and\ \bibinfo {author} {\bibfnamefont {J.-G.}\ \bibnamefont {Shin}},\
  }\bibfield  {title} {\bibinfo {title} {Two peaks in the momentum distribution
  of bosons in a weakly frustrated two-leg optical ladder},\ }\href@noop {}
  {\bibfield  {journal} {\bibinfo  {journal} {Physical Review A}\ }\textbf
  {\bibinfo {volume} {83}},\ \bibinfo {pages} {055602} (\bibinfo {year}
  {2011})}\BibitemShut {NoStop}%
\bibitem [{\citenamefont {Piraud}\ \emph {et~al.}(2014)\citenamefont {Piraud},
  \citenamefont {Cai}, \citenamefont {McCulloch},\ and\ \citenamefont
  {Schollw\"ock}}]{Piraud2014}%
  \BibitemOpen
  \bibfield  {author} {\bibinfo {author} {\bibfnamefont {M.}~\bibnamefont
  {Piraud}}, \bibinfo {author} {\bibfnamefont {Z.}~\bibnamefont {Cai}},
  \bibinfo {author} {\bibfnamefont {I.~P.}\ \bibnamefont {McCulloch}},\ and\
  \bibinfo {author} {\bibfnamefont {U.}~\bibnamefont {Schollw\"ock}},\
  }\bibfield  {title} {\bibinfo {title} {Quantum magnetism of bosons with
  synthetic gauge fields in one-dimensional optical lattices: a density-matrix
  renormalization group study},\ }\href@noop {} {\bibfield  {journal} {\bibinfo
   {journal} {Phys. Rev. A}\ }\textbf {\bibinfo {volume} {89}},\ \bibinfo
  {pages} {063618} (\bibinfo {year} {2014})}\BibitemShut {NoStop}%
\bibitem [{\citenamefont {Piraud}\ \emph
  {et~al.}(2015{\natexlab{a}})\citenamefont {Piraud}, \citenamefont
  {Heidrich-Meisner}, \citenamefont {McCulloch}, \citenamefont {Greschner},
  \citenamefont {Vekua},\ and\ \citenamefont {Schollw\"ock}}]{piraud2014b}%
  \BibitemOpen
  \bibfield  {author} {\bibinfo {author} {\bibfnamefont {M.}~\bibnamefont
  {Piraud}}, \bibinfo {author} {\bibfnamefont {F.}~\bibnamefont
  {Heidrich-Meisner}}, \bibinfo {author} {\bibfnamefont {I.~P.}\ \bibnamefont
  {McCulloch}}, \bibinfo {author} {\bibfnamefont {S.}~\bibnamefont
  {Greschner}}, \bibinfo {author} {\bibfnamefont {T.}~\bibnamefont {Vekua}},\
  and\ \bibinfo {author} {\bibfnamefont {U.}~\bibnamefont {Schollw\"ock}},\
  }\bibfield  {title} {\bibinfo {title} {Vortex and meissner phases of strongly
  interacting bosons on a two-leg ladder},\ }\href
  {https://doi.org/10.1103/PhysRevB.91.140406} {\bibfield  {journal} {\bibinfo
  {journal} {Physical Review B}\ }\textbf {\bibinfo {volume} {91}},\ \bibinfo
  {pages} {140406} (\bibinfo {year} {2015}{\natexlab{a}})}\BibitemShut
  {NoStop}%
\bibitem [{\citenamefont {Piraud}\ \emph
  {et~al.}(2015{\natexlab{b}})\citenamefont {Piraud}, \citenamefont
  {Heidrich-Meisner}, \citenamefont {McCulloch}, \citenamefont {Greschner},
  \citenamefont {Vekua},\ and\ \citenamefont {Schollw\"ock}}]{Piraud2015}%
  \BibitemOpen
  \bibfield  {author} {\bibinfo {author} {\bibfnamefont {M.}~\bibnamefont
  {Piraud}}, \bibinfo {author} {\bibfnamefont {F.}~\bibnamefont
  {Heidrich-Meisner}}, \bibinfo {author} {\bibfnamefont {I.~P.}\ \bibnamefont
  {McCulloch}}, \bibinfo {author} {\bibfnamefont {S.}~\bibnamefont
  {Greschner}}, \bibinfo {author} {\bibfnamefont {T.}~\bibnamefont {Vekua}},\
  and\ \bibinfo {author} {\bibfnamefont {U.}~\bibnamefont {Schollw\"ock}},\
  }\bibfield  {title} {\bibinfo {title} {Vortex and meissner phases of strongly
  interacting bosons on a two-leg ladder},\ }\href
  {https://doi.org/10.1103/PhysRevB.91.140406} {\bibfield  {journal} {\bibinfo
  {journal} {Phys. Rev. B}\ }\textbf {\bibinfo {volume} {91}},\ \bibinfo
  {pages} {140406(R)} (\bibinfo {year} {2015}{\natexlab{b}})}\BibitemShut
  {NoStop}%
\bibitem [{\citenamefont {Di~Dio}\ \emph {et~al.}(2015)\citenamefont {Di~Dio},
  \citenamefont {De~Palo}, \citenamefont {Orignac}, \citenamefont {Citro},\
  and\ \citenamefont {Chiofalo}}]{DiDio2015}%
  \BibitemOpen
  \bibfield  {author} {\bibinfo {author} {\bibfnamefont {M.}~\bibnamefont
  {Di~Dio}}, \bibinfo {author} {\bibfnamefont {S.}~\bibnamefont {De~Palo}},
  \bibinfo {author} {\bibfnamefont {E.}~\bibnamefont {Orignac}}, \bibinfo
  {author} {\bibfnamefont {R.}~\bibnamefont {Citro}},\ and\ \bibinfo {author}
  {\bibfnamefont {M.-L.}\ \bibnamefont {Chiofalo}},\ }\bibfield  {title}
  {\bibinfo {title} {Persisting meissner state and incommensurate phases of
  hard-core boson ladders in a flux},\ }\href@noop {} {\bibfield  {journal}
  {\bibinfo  {journal} {Phys. Rev. B}\ }\textbf {\bibinfo {volume} {92}},\
  \bibinfo {pages} {060506} (\bibinfo {year} {2015})}\BibitemShut {NoStop}%
\bibitem [{\citenamefont {Greschner}\ \emph {et~al.}(2015)\citenamefont
  {Greschner}, \citenamefont {Piraud}, \citenamefont {Heidrich-Meisner},
  \citenamefont {McCulloch}, \citenamefont {Schollw{\"o}ck},\ and\
  \citenamefont {Vekua}}]{greschner2015}%
  \BibitemOpen
  \bibfield  {author} {\bibinfo {author} {\bibfnamefont {S.}~\bibnamefont
  {Greschner}}, \bibinfo {author} {\bibfnamefont {M.}~\bibnamefont {Piraud}},
  \bibinfo {author} {\bibfnamefont {F.}~\bibnamefont {Heidrich-Meisner}},
  \bibinfo {author} {\bibfnamefont {I.}~\bibnamefont {McCulloch}}, \bibinfo
  {author} {\bibfnamefont {U.}~\bibnamefont {Schollw{\"o}ck}},\ and\ \bibinfo
  {author} {\bibfnamefont {T.}~\bibnamefont {Vekua}},\ }\bibfield  {title}
  {\bibinfo {title} {Spontaneous increase of magnetic flux and chiral-current
  reversal in bosonic ladders: Swimming against the tide},\ }\href@noop {}
  {\bibfield  {journal} {\bibinfo  {journal} {Phys. Rev. Lett.}\ }\textbf
  {\bibinfo {volume} {115}},\ \bibinfo {pages} {190402} (\bibinfo {year}
  {2015})}\BibitemShut {NoStop}%
\bibitem [{\citenamefont {Natu}(2015)}]{Natu2015}%
  \BibitemOpen
  \bibfield  {author} {\bibinfo {author} {\bibfnamefont {S.}~\bibnamefont
  {Natu}},\ }\bibfield  {title} {\bibinfo {title} {Bosons with long range
  interactions on two-leg ladders in artificial magnetic fields stefan s.
  natu},\ }\href@noop {} {\bibfield  {journal} {\bibinfo  {journal} {Phys. Rev.
  A}\ }\textbf {\bibinfo {volume} {92}},\ \bibinfo {pages} {053623} (\bibinfo
  {year} {2015})}\BibitemShut {NoStop}%
\bibitem [{\citenamefont {Greschner}\ \emph {et~al.}(6 12)\citenamefont
  {Greschner}, \citenamefont {Piraud}, \citenamefont {Heidrich-Meisner},
  \citenamefont {McCulloch}, \citenamefont {Schollw\"ock},\ and\ \citenamefont
  {Vekua}}]{greschner2016}%
  \BibitemOpen
  \bibfield  {author} {\bibinfo {author} {\bibfnamefont {S.}~\bibnamefont
  {Greschner}}, \bibinfo {author} {\bibfnamefont {M.}~\bibnamefont {Piraud}},
  \bibinfo {author} {\bibfnamefont {F.}~\bibnamefont {Heidrich-Meisner}},
  \bibinfo {author} {\bibfnamefont {I.~P.}\ \bibnamefont {McCulloch}}, \bibinfo
  {author} {\bibfnamefont {U.}~\bibnamefont {Schollw\"ock}},\ and\ \bibinfo
  {author} {\bibfnamefont {T.}~\bibnamefont {Vekua}},\ }\bibfield  {title}
  {\bibinfo {title} {Symmetry-broken states in a system of interacting bosons
  on a two-leg ladder with a uniform abelian gauge field},\ }\href@noop {}
  {\bibfield  {journal} {\bibinfo  {journal} {Phys. Rev. A}\ }\textbf {\bibinfo
  {volume} {94}},\ \bibinfo {pages} {063628} (\bibinfo {year}
  {2016-12})}\BibitemShut {NoStop}%
\bibitem [{\citenamefont {Greschner}\ and\ \citenamefont {Vekua}(7
  08)}]{greschner2017}%
  \BibitemOpen
  \bibfield  {author} {\bibinfo {author} {\bibfnamefont {S.}~\bibnamefont
  {Greschner}}\ and\ \bibinfo {author} {\bibfnamefont {T.}~\bibnamefont
  {Vekua}},\ }\bibfield  {title} {\bibinfo {title} {Vortex-hole duality: A
  unified picture of weak- and strong-coupling regimes of bosonic ladders with
  flux},\ }\href {https://doi.org/10.1103/PhysRevLett.119.073401} {\bibfield
  {journal} {\bibinfo  {journal} {Phys. Rev. Lett.}\ }\textbf {\bibinfo
  {volume} {119}},\ \bibinfo {pages} {073401} (\bibinfo {year}
  {2017-08})}\BibitemShut {NoStop}%
\bibitem [{\citenamefont {Orignac}\ \emph {et~al.}(7 07)\citenamefont
  {Orignac}, \citenamefont {Citro}, \citenamefont {Di~Dio},\ and\ \citenamefont
  {De~Palo}}]{Orignac2017}%
  \BibitemOpen
  \bibfield  {author} {\bibinfo {author} {\bibfnamefont {E.}~\bibnamefont
  {Orignac}}, \bibinfo {author} {\bibfnamefont {R.}~\bibnamefont {Citro}},
  \bibinfo {author} {\bibfnamefont {M.}~\bibnamefont {Di~Dio}},\ and\ \bibinfo
  {author} {\bibfnamefont {S.}~\bibnamefont {De~Palo}},\ }\bibfield  {title}
  {\bibinfo {title} {Vortex lattice melting in a boson ladder in an artificial
  gauge field},\ }\href {https://doi.org/10.1103/PhysRevB.96.014518} {\bibfield
   {journal} {\bibinfo  {journal} {Phys. Rev. B}\ }\textbf {\bibinfo {volume}
  {96}},\ \bibinfo {pages} {014518} (\bibinfo {year} {2017-07})}\BibitemShut
  {NoStop}%
\bibitem [{\citenamefont {Citro}\ \emph {et~al.}(2018)\citenamefont {Citro},
  \citenamefont {De~Palo}, \citenamefont {Di~Dio},\ and\ \citenamefont
  {Orignac}}]{Citro2018}%
  \BibitemOpen
  \bibfield  {author} {\bibinfo {author} {\bibfnamefont {R.}~\bibnamefont
  {Citro}}, \bibinfo {author} {\bibfnamefont {S.}~\bibnamefont {De~Palo}},
  \bibinfo {author} {\bibfnamefont {M.}~\bibnamefont {Di~Dio}},\ and\ \bibinfo
  {author} {\bibfnamefont {E.}~\bibnamefont {Orignac}},\ }\bibfield  {title}
  {\bibinfo {title} {Quantum phase transitions of a two-leg bosonic ladder in
  an artificial gauge field},\ }\href@noop {} {\bibfield  {journal} {\bibinfo
  {journal} {Physical Review B}\ }\textbf {\bibinfo {volume} {97}},\ \bibinfo
  {pages} {174523} (\bibinfo {year} {2018})}\BibitemShut {NoStop}%
\bibitem [{\citenamefont {Maeda}\ \emph {et~al.}(2007)\citenamefont {Maeda},
  \citenamefont {Hotta},\ and\ \citenamefont
  {Oshikawa}}]{Maeda_spinchain_magnetization}%
  \BibitemOpen
  \bibfield  {author} {\bibinfo {author} {\bibfnamefont {Y.}~\bibnamefont
  {Maeda}}, \bibinfo {author} {\bibfnamefont {C.}~\bibnamefont {Hotta}},\ and\
  \bibinfo {author} {\bibfnamefont {M.}~\bibnamefont {Oshikawa}},\ }\bibfield
  {title} {\bibinfo {title} {Universal temperature dependence of the
  magnetization of gapped spin chains},\ }\href@noop {} {\bibfield  {journal}
  {\bibinfo  {journal} {Physical Review Letters}\ }\textbf {\bibinfo {volume}
  {99}},\ \bibinfo {pages} {057205} (\bibinfo {year} {2007})}\BibitemShut
  {NoStop}%
\bibitem [{\citenamefont {Buser}\ \emph {et~al.}(2019)\citenamefont {Buser},
  \citenamefont {Heidrich-Meisner},\ and\ \citenamefont
  {Schollwöck}}]{buser_finite-temperature_2019}%
  \BibitemOpen
  \bibfield  {author} {\bibinfo {author} {\bibfnamefont {M.}~\bibnamefont
  {Buser}}, \bibinfo {author} {\bibfnamefont {F.}~\bibnamefont
  {Heidrich-Meisner}},\ and\ \bibinfo {author} {\bibfnamefont {U.}~\bibnamefont
  {Schollwöck}},\ }\bibfield  {title} {\bibinfo {title} {Finite-temperature
  properties of interacting bosons on a two-leg flux ladder},\ }\href
  {https://doi.org/10.1103/PhysRevA.99.053601} {\bibfield  {journal} {\bibinfo
  {journal} {Physical Review A}\ }\textbf {\bibinfo {volume} {99}},\ \bibinfo
  {pages} {053601} (\bibinfo {year} {2019})},\ \bibinfo {note} {arXiv:
  1901.07083}\BibitemShut {NoStop}%
\bibitem [{\citenamefont {Atala}\ \emph {et~al.}(2014)\citenamefont {Atala},
  \citenamefont {Aidelsburger}, \citenamefont {Lohse}, \citenamefont
  {Barreiro}, \citenamefont {Paredes},\ and\ \citenamefont
  {Bloch}}]{atala_2014}%
  \BibitemOpen
  \bibfield  {author} {\bibinfo {author} {\bibfnamefont {M.}~\bibnamefont
  {Atala}}, \bibinfo {author} {\bibfnamefont {M.}~\bibnamefont {Aidelsburger}},
  \bibinfo {author} {\bibfnamefont {M.}~\bibnamefont {Lohse}}, \bibinfo
  {author} {\bibfnamefont {J.~T.}\ \bibnamefont {Barreiro}}, \bibinfo {author}
  {\bibfnamefont {B.}~\bibnamefont {Paredes}},\ and\ \bibinfo {author}
  {\bibfnamefont {I.}~\bibnamefont {Bloch}},\ }\bibfield  {title} {\bibinfo
  {title} {Observation of chiral currents with ultracold atoms in bosonic
  ladders},\ }\href@noop {} {\bibfield  {journal} {\bibinfo  {journal} {Nature
  Physics}\ }\textbf {\bibinfo {volume} {10}},\ \bibinfo {pages} {588}
  (\bibinfo {year} {2014})}\BibitemShut {NoStop}%
\bibitem [{\citenamefont {Petrescu}\ and\ \citenamefont
  {Le~Hur}(2015)}]{Petrescu2015}%
  \BibitemOpen
  \bibfield  {author} {\bibinfo {author} {\bibfnamefont {A.}~\bibnamefont
  {Petrescu}}\ and\ \bibinfo {author} {\bibfnamefont {K.}~\bibnamefont
  {Le~Hur}},\ }\bibfield  {title} {\bibinfo {title} {Chiral mott insulators,
  meissner effect, and laughlin states in quantum ladders},\ }\href@noop {}
  {\bibfield  {journal} {\bibinfo  {journal} {Phys. Rev. B}\ }\textbf {\bibinfo
  {volume} {91}},\ \bibinfo {pages} {054520} (\bibinfo {year}
  {2015})}\BibitemShut {NoStop}%
\bibitem [{\citenamefont {Cornfeld}\ and\ \citenamefont
  {Sela}(2015)}]{Cornfeld2015}%
  \BibitemOpen
  \bibfield  {author} {\bibinfo {author} {\bibfnamefont {E.}~\bibnamefont
  {Cornfeld}}\ and\ \bibinfo {author} {\bibfnamefont {E.}~\bibnamefont
  {Sela}},\ }\bibfield  {title} {\bibinfo {title} {Chiral currents in
  one-dimensional fractional quantum hall states},\ }\href@noop {} {\bibfield
  {journal} {\bibinfo  {journal} {Phys. Rev. B}\ }\textbf {\bibinfo {volume}
  {92}},\ \bibinfo {pages} {115446} (\bibinfo {year} {2015})}\BibitemShut
  {NoStop}%
\bibitem [{\citenamefont {Petrescu}\ \emph {et~al.}(7 07)\citenamefont
  {Petrescu}, \citenamefont {Piraud}, \citenamefont {Roux}, \citenamefont
  {McCulloch},\ and\ \citenamefont {Le~Hur}}]{Petrescu2017}%
  \BibitemOpen
  \bibfield  {author} {\bibinfo {author} {\bibfnamefont {A.}~\bibnamefont
  {Petrescu}}, \bibinfo {author} {\bibfnamefont {M.}~\bibnamefont {Piraud}},
  \bibinfo {author} {\bibfnamefont {G.}~\bibnamefont {Roux}}, \bibinfo {author}
  {\bibfnamefont {I.~P.}\ \bibnamefont {McCulloch}},\ and\ \bibinfo {author}
  {\bibfnamefont {K.}~\bibnamefont {Le~Hur}},\ }\bibfield  {title} {\bibinfo
  {title} {Precursor of the laughlin state of hard-core bosons on a two-leg
  ladder},\ }\href {https://doi.org/10.1103/PhysRevB.96.014524} {\bibfield
  {journal} {\bibinfo  {journal} {Phys. Rev. B}\ }\textbf {\bibinfo {volume}
  {96}},\ \bibinfo {pages} {014524} (\bibinfo {year} {2017-07})}\BibitemShut
  {NoStop}%
\bibitem [{\citenamefont {Strinati}\ \emph {et~al.}(2017)\citenamefont
  {Strinati}, \citenamefont {Cornfeld}, \citenamefont {Rossini}, \citenamefont
  {Barbarino}, \citenamefont {Dalmonte}, \citenamefont {Fazio}, \citenamefont
  {Sela},\ and\ \citenamefont {Mazza}}]{Strinati2017}%
  \BibitemOpen
  \bibfield  {author} {\bibinfo {author} {\bibfnamefont {M.~C.}\ \bibnamefont
  {Strinati}}, \bibinfo {author} {\bibfnamefont {E.}~\bibnamefont {Cornfeld}},
  \bibinfo {author} {\bibfnamefont {D.}~\bibnamefont {Rossini}}, \bibinfo
  {author} {\bibfnamefont {S.}~\bibnamefont {Barbarino}}, \bibinfo {author}
  {\bibfnamefont {M.}~\bibnamefont {Dalmonte}}, \bibinfo {author}
  {\bibfnamefont {R.}~\bibnamefont {Fazio}}, \bibinfo {author} {\bibfnamefont
  {E.}~\bibnamefont {Sela}},\ and\ \bibinfo {author} {\bibfnamefont
  {L.}~\bibnamefont {Mazza}},\ }\bibfield  {title} {\bibinfo {title}
  {Laughlin-like states in bosonic and fermionic atomic synthetic ladders},\
  }\href@noop {} {\bibfield  {journal} {\bibinfo  {journal} {Physical Review
  X}\ }\textbf {\bibinfo {volume} {7}},\ \bibinfo {pages} {021033} (\bibinfo
  {year} {2017})}\BibitemShut {NoStop}%
\bibitem [{\citenamefont {Strinati}\ \emph {et~al.}(2019)\citenamefont
  {Strinati}, \citenamefont {Sahoo}, \citenamefont {Shtengel},\ and\
  \citenamefont {Sela}}]{strinati2019}%
  \BibitemOpen
  \bibfield  {author} {\bibinfo {author} {\bibfnamefont {M.~C.}\ \bibnamefont
  {Strinati}}, \bibinfo {author} {\bibfnamefont {S.}~\bibnamefont {Sahoo}},
  \bibinfo {author} {\bibfnamefont {K.}~\bibnamefont {Shtengel}},\ and\
  \bibinfo {author} {\bibfnamefont {E.}~\bibnamefont {Sela}},\ }\bibfield
  {title} {\bibinfo {title} {Pretopological fractional excitations in the
  two-leg flux ladder},\ }\href@noop {} {\bibfield  {journal} {\bibinfo
  {journal} {Physical Review B}\ }\textbf {\bibinfo {volume} {99}},\ \bibinfo
  {pages} {245101} (\bibinfo {year} {2019})}\BibitemShut {NoStop}%
\bibitem [{\citenamefont {Haller}\ \emph {et~al.}(2020)\citenamefont {Haller},
  \citenamefont {Matsoukas-Roubeas}, \citenamefont {Pan}, \citenamefont
  {Rizzi},\ and\ \citenamefont {Burrello}}]{haller_2020}%
  \BibitemOpen
  \bibfield  {author} {\bibinfo {author} {\bibfnamefont {A.}~\bibnamefont
  {Haller}}, \bibinfo {author} {\bibfnamefont {A.~S.}\ \bibnamefont
  {Matsoukas-Roubeas}}, \bibinfo {author} {\bibfnamefont {Y.}~\bibnamefont
  {Pan}}, \bibinfo {author} {\bibfnamefont {M.}~\bibnamefont {Rizzi}},\ and\
  \bibinfo {author} {\bibfnamefont {M.}~\bibnamefont {Burrello}},\ }\bibfield
  {title} {\bibinfo {title} {Exploring helical phases of matter in bosonic
  ladders},\ }\href {https://doi.org/10.1103/PhysRevResearch.2.043433}
  {\bibfield  {journal} {\bibinfo  {journal} {Phys. Rev. Research}\ }\textbf
  {\bibinfo {volume} {2}},\ \bibinfo {pages} {043433} (\bibinfo {year}
  {2020})},\ \bibinfo {note} {arXiv: 2010.02740}\BibitemShut {NoStop}%
\bibitem [{\citenamefont {Giamarchi}(2004)}]{giamarchi_book_1d}%
  \BibitemOpen
  \bibfield  {author} {\bibinfo {author} {\bibfnamefont {T.}~\bibnamefont
  {Giamarchi}},\ }\href@noop {} {\emph {\bibinfo {title} {Quantum Physics in
  One Dimension}}},\ \bibinfo {series} {International series of monographs on
  physics}, Vol.\ \bibinfo {volume} {121}\ (\bibinfo  {publisher} {Oxford
  University Press},\ \bibinfo {address} {Oxford},\ \bibinfo {year}
  {2004})\BibitemShut {NoStop}%
\bibitem [{\citenamefont {Schick}(1968)}]{schick_flux_1968}%
  \BibitemOpen
  \bibfield  {author} {\bibinfo {author} {\bibfnamefont {M.}~\bibnamefont
  {Schick}},\ }\bibfield  {title} {\bibinfo {title} {Flux {Quantization} in a
  {One}-{Dimensional} {Model}},\ }\href
  {https://doi.org/10.1103/PhysRev.166.404} {\bibfield  {journal} {\bibinfo
  {journal} {Phys. Rev.}\ }\textbf {\bibinfo {volume} {166}},\ \bibinfo {pages}
  {404} (\bibinfo {year} {1968})}\BibitemShut {NoStop}%
\bibitem [{\citenamefont {Haldane}(1981{\natexlab{b}})}]{haldane_bosonisation}%
  \BibitemOpen
  \bibfield  {author} {\bibinfo {author} {\bibfnamefont {F.~D.~M.}\
  \bibnamefont {Haldane}},\ }\href@noop {} {\bibfield  {journal} {\bibinfo
  {journal} {J. Phys. C}\ }\textbf {\bibinfo {volume} {14}},\ \bibinfo {pages}
  {2585} (\bibinfo {year} {1981}{\natexlab{b}})}\BibitemShut {NoStop}%
\bibitem [{\citenamefont {Lopatin}\ \emph {et~al.}(2001)\citenamefont
  {Lopatin}, \citenamefont {Georges},\ and\ \citenamefont
  {Giamarchi}}]{lopatin_q1d_magnetooptical}%
  \BibitemOpen
  \bibfield  {author} {\bibinfo {author} {\bibfnamefont {A.}~\bibnamefont
  {Lopatin}}, \bibinfo {author} {\bibfnamefont {A.}~\bibnamefont {Georges}},\
  and\ \bibinfo {author} {\bibfnamefont {T.}~\bibnamefont {Giamarchi}},\
  }\bibfield  {title} {\bibinfo {title} {Hall effect and inter-chain
  magneto-optical properties of coupled luttinger liquids},\ }\href@noop {}
  {\bibfield  {journal} {\bibinfo  {journal} {Physical Review B}\ }\textbf
  {\bibinfo {volume} {63}},\ \bibinfo {pages} {75109} (\bibinfo {year}
  {2001})},\ \Eprint {https://arxiv.org/abs/cond-mat/0008066}
  {cond-mat/0008066} \BibitemShut {NoStop}%
\bibitem [{\citenamefont {Kohn}(1964)}]{kohn_stiffness}%
  \BibitemOpen
  \bibfield  {author} {\bibinfo {author} {\bibfnamefont {W.}~\bibnamefont
  {Kohn}},\ }\href@noop {} {\bibfield  {journal} {\bibinfo  {journal} {Physical
  Review}\ }\textbf {\bibinfo {volume} {133}},\ \bibinfo {pages} {A171}
  (\bibinfo {year} {1964})}\BibitemShut {NoStop}%
\bibitem [{\citenamefont {Schüttelkopf}\ \emph {et~al.}(2024)\citenamefont
  {Schüttelkopf}, \citenamefont {Tajik}, \citenamefont {Bazhan}, \citenamefont
  {Cataldini}, \citenamefont {Ji}, \citenamefont {Schmiedmayer},\ and\
  \citenamefont {Møller}}]{schuttelkopf_drude_weight_2024}%
  \BibitemOpen
  \bibfield  {author} {\bibinfo {author} {\bibfnamefont {P.}~\bibnamefont
  {Schüttelkopf}}, \bibinfo {author} {\bibfnamefont {M.}~\bibnamefont
  {Tajik}}, \bibinfo {author} {\bibfnamefont {N.}~\bibnamefont {Bazhan}},
  \bibinfo {author} {\bibfnamefont {F.}~\bibnamefont {Cataldini}}, \bibinfo
  {author} {\bibfnamefont {S.-C.}\ \bibnamefont {Ji}}, \bibinfo {author}
  {\bibfnamefont {J.}~\bibnamefont {Schmiedmayer}},\ and\ \bibinfo {author}
  {\bibfnamefont {F.}~\bibnamefont {Møller}},\ }\href
  {https://doi.org/10.48550/arXiv.2406.17569} {\bibinfo {title} {Characterising
  transport in a quantum gas by measuring {Drude} weights}},\ \bibinfo
  {howpublished} {arXiv:2406.17569 [cond-mat]} (\bibinfo {year}
  {2024})\BibitemShut {NoStop}%
\bibitem [{\citenamefont {Donohue}\ and\ \citenamefont
  {Giamarchi}(2001)}]{donohue_commensurate_bosonic_ladder}%
  \BibitemOpen
  \bibfield  {author} {\bibinfo {author} {\bibfnamefont {P.}~\bibnamefont
  {Donohue}}\ and\ \bibinfo {author} {\bibfnamefont {T.}~\bibnamefont
  {Giamarchi}},\ }\bibfield  {title} {\bibinfo {title} {Mott-superfluid
  transition in bosonic ladders},\ }\href@noop {} {\bibfield  {journal}
  {\bibinfo  {journal} {Physical Review B}\ }\textbf {\bibinfo {volume} {63}},\
  \bibinfo {pages} {180508(R)} (\bibinfo {year} {2001})}\BibitemShut {NoStop}%
\bibitem [{\citenamefont {Cr\'epin}\ \emph {et~al.}(1 08)\citenamefont
  {Cr\'epin}, \citenamefont {Laflorencie}, \citenamefont {Roux},\ and\
  \citenamefont {Simon}}]{Crepin2011}%
  \BibitemOpen
  \bibfield  {author} {\bibinfo {author} {\bibfnamefont {F.}~\bibnamefont
  {Cr\'epin}}, \bibinfo {author} {\bibfnamefont {N.}~\bibnamefont
  {Laflorencie}}, \bibinfo {author} {\bibfnamefont {G.}~\bibnamefont {Roux}},\
  and\ \bibinfo {author} {\bibfnamefont {P.}~\bibnamefont {Simon}},\ }\bibfield
   {title} {\bibinfo {title} {Phase diagram of hard-core bosons on clean and
  disordered two-leg ladders: Mott insulator\char21{}luttinger
  liquid\char21{}bose glass},\ }\href
  {https://doi.org/10.1103/PhysRevB.84.054517} {\bibfield  {journal} {\bibinfo
  {journal} {Phys. Rev. B}\ }\textbf {\bibinfo {volume} {84}},\ \bibinfo
  {pages} {054517} (\bibinfo {year} {2011-08})}\BibitemShut {NoStop}%
\bibitem [{\citenamefont {Kolley}\ \emph {et~al.}(2015)\citenamefont {Kolley},
  \citenamefont {Piraud}, \citenamefont {McCulloch}, \citenamefont
  {Schollw{\"o}ck},\ and\ \citenamefont {Heidrich-Meisner}}]{Kolley2015}%
  \BibitemOpen
  \bibfield  {author} {\bibinfo {author} {\bibfnamefont {F.}~\bibnamefont
  {Kolley}}, \bibinfo {author} {\bibfnamefont {M.}~\bibnamefont {Piraud}},
  \bibinfo {author} {\bibfnamefont {I.}~\bibnamefont {McCulloch}}, \bibinfo
  {author} {\bibfnamefont {U.}~\bibnamefont {Schollw{\"o}ck}},\ and\ \bibinfo
  {author} {\bibfnamefont {F.}~\bibnamefont {Heidrich-Meisner}},\ }\bibfield
  {title} {\bibinfo {title} {Strongly interacting bosons on a three-leg ladder
  in the presence of a homogeneous flux},\ }\href@noop {} {\bibfield  {journal}
  {\bibinfo  {journal} {New J. Phys.}\ }\textbf {\bibinfo {volume} {17}},\
  \bibinfo {pages} {092001} (\bibinfo {year} {2015})}\BibitemShut {NoStop}%
\bibitem [{\citenamefont {Citro}\ \emph {et~al.}(2000)\citenamefont {Citro},
  \citenamefont {Orignac}, \citenamefont {Andrei}, \citenamefont {Itoi},\ and\
  \citenamefont {Qin}}]{citro2000}%
  \BibitemOpen
  \bibfield  {author} {\bibinfo {author} {\bibfnamefont {R.}~\bibnamefont
  {Citro}}, \bibinfo {author} {\bibfnamefont {E.}~\bibnamefont {Orignac}},
  \bibinfo {author} {\bibfnamefont {N.}~\bibnamefont {Andrei}}, \bibinfo
  {author} {\bibfnamefont {C.}~\bibnamefont {Itoi}},\ and\ \bibinfo {author}
  {\bibfnamefont {S.}~\bibnamefont {Qin}},\ }\bibfield  {title} {\bibinfo
  {title} {Effective theory of magnetization plateaus in a three-leg ladder
  with periodic boundary conditions},\ }\href@noop {} {\bibfield  {journal}
  {\bibinfo  {journal} {Journal of Physics: Condensed Matter}\ }\textbf
  {\bibinfo {volume} {12}},\ \bibinfo {pages} {3041} (\bibinfo {year}
  {2000})}\BibitemShut {NoStop}%
\bibitem [{\citenamefont {Bouchoule}\ \emph {et~al.}(2025)\citenamefont
  {Bouchoule}, \citenamefont {Citro}, \citenamefont {Duty}, \citenamefont
  {Giamarchi}, \citenamefont {Hulet}, \citenamefont {Klanjsek}, \citenamefont
  {Orignac},\ and\ \citenamefont {Weber}}]{tll_review_2025}%
  \BibitemOpen
  \bibfield  {author} {\bibinfo {author} {\bibfnamefont {I.}~\bibnamefont
  {Bouchoule}}, \bibinfo {author} {\bibfnamefont {R.}~\bibnamefont {Citro}},
  \bibinfo {author} {\bibfnamefont {T.}~\bibnamefont {Duty}}, \bibinfo {author}
  {\bibfnamefont {T.}~\bibnamefont {Giamarchi}}, \bibinfo {author}
  {\bibfnamefont {R.~G.}\ \bibnamefont {Hulet}}, \bibinfo {author}
  {\bibfnamefont {M.}~\bibnamefont {Klanjsek}}, \bibinfo {author}
  {\bibfnamefont {E.}~\bibnamefont {Orignac}},\ and\ \bibinfo {author}
  {\bibfnamefont {B.}~\bibnamefont {Weber}},\ }\bibfield  {title} {\bibinfo
  {title} {A glance to luttinger liquid and its platforms},\ }\href
  {https://doi.org/10.1038/s42254-025-00866-w} {\bibfield  {journal} {\bibinfo
  {journal} {Nature Review Physics}\ }\textbf {\bibinfo {volume} {7}},\
  \bibinfo {pages} {565} (\bibinfo {year} {2025})},\ \Eprint
  {https://arxiv.org/abs/arXiv:2501.12097} {arXiv:2501.12097} \BibitemShut
  {NoStop}%
\bibitem [{\citenamefont {Japaridze}\ and\ \citenamefont
  {Nersesyan}(1978)}]{japaridze_cic_transition}%
  \BibitemOpen
  \bibfield  {author} {\bibinfo {author} {\bibfnamefont {G.~I.}\ \bibnamefont
  {Japaridze}}\ and\ \bibinfo {author} {\bibfnamefont {A.~A.}\ \bibnamefont
  {Nersesyan}},\ }\bibfield  {title} {\bibinfo {title} {?},\ }\href@noop {}
  {\bibfield  {journal} {\bibinfo  {journal} {Journal of Experimental and
  Theoretical Physics Letters}\ }\textbf {\bibinfo {volume} {27}},\ \bibinfo
  {pages} {334} (\bibinfo {year} {1978})}\BibitemShut {NoStop}%
\bibitem [{\citenamefont {Pokrovsky}\ and\ \citenamefont
  {Talapov}(1979)}]{pokrovsky_talapov_prl}%
  \BibitemOpen
  \bibfield  {author} {\bibinfo {author} {\bibfnamefont {V.~L.}\ \bibnamefont
  {Pokrovsky}}\ and\ \bibinfo {author} {\bibfnamefont {A.~L.}\ \bibnamefont
  {Talapov}},\ }\bibfield  {title} {\bibinfo {title} {{No Title}},\ }\href@noop
  {} {\bibfield  {journal} {\bibinfo  {journal} {Physical Review Letters}\
  }\textbf {\bibinfo {volume} {42}},\ \bibinfo {pages} {65} (\bibinfo {year}
  {1979})}\BibitemShut {NoStop}%
\bibitem [{\citenamefont {Matveev}(2013)}]{matveev_equilibration_2013}%
  \BibitemOpen
  \bibfield  {author} {\bibinfo {author} {\bibfnamefont {K.~A.}\ \bibnamefont
  {Matveev}},\ }\bibfield  {title} {\bibinfo {title} {Equilibration of a
  one-dimensional quantum liquid},\ }\href
  {https://doi.org/10.1134/S1063776113110137} {\bibfield  {journal} {\bibinfo
  {journal} {Journal of Experimental and Theoretical Physics}\ }\textbf
  {\bibinfo {volume} {117}},\ \bibinfo {pages} {508} (\bibinfo {year}
  {2013})},\ \bibinfo {note} {arXiv:1304.6012 [cond-mat]}\BibitemShut {NoStop}%
\bibitem [{\citenamefont {Citro}\ \emph {et~al.}(2025)\citenamefont {Citro},
  \citenamefont {Giamarchi},\ and\ \citenamefont {Orignac}}]{citro_hall_2025}%
  \BibitemOpen
  \bibfield  {author} {\bibinfo {author} {\bibfnamefont {R.}~\bibnamefont
  {Citro}}, \bibinfo {author} {\bibfnamefont {T.}~\bibnamefont {Giamarchi}},\
  and\ \bibinfo {author} {\bibfnamefont {E.}~\bibnamefont {Orignac}},\
  }\bibfield  {title} {\bibinfo {title} {Hall {Response} in {Interacting}
  {Bosonic} and {Fermionic} {Ladders}},\ }\href
  {https://doi.org/10.1103/PhysRevLett.134.056501} {\bibfield  {journal}
  {\bibinfo  {journal} {Physical Review Letters}\ }\textbf {\bibinfo {volume}
  {134}},\ \bibinfo {pages} {056501} (\bibinfo {year} {2025})},\ \bibinfo
  {note} {arXiv:2404.16973 [cond-mat]}\BibitemShut {NoStop}%
\bibitem [{\citenamefont {Zotos}\ \emph {et~al.}(2001)\citenamefont {Zotos},
  \citenamefont {Naef}, \citenamefont {Long},\ and\ \citenamefont
  {Prelovšek}}]{zotos_drude_2001}%
  \BibitemOpen
  \bibfield  {author} {\bibinfo {author} {\bibfnamefont {X.}~\bibnamefont
  {Zotos}}, \bibinfo {author} {\bibfnamefont {F.}~\bibnamefont {Naef}},
  \bibinfo {author} {\bibfnamefont {M.}~\bibnamefont {Long}},\ and\ \bibinfo
  {author} {\bibfnamefont {P.}~\bibnamefont {Prelovšek}},\ }\bibfield  {title}
  {\bibinfo {title} {Drude {Weight}, {Integrable} {Systems} and the {Reactive}
  {Hall} {Constant}},\ }in\ \href
  {https://doi.org/10.1007/978-94-010-0771-9_28} {\emph {\bibinfo {booktitle}
  {Open {Problems} in {Strongly} {Correlated} {Electron} {Systems}}}},\
  \bibinfo {series} {{NATO} {Science} {Series} {II}: {Mathematics}, {Physics}
  and {Chemistry}}, Vol.~\bibinfo {volume} {15},\ \bibinfo {editor} {edited by\
  \bibinfo {editor} {\bibfnamefont {J.}~\bibnamefont {Bonča}}, \bibinfo
  {editor} {\bibfnamefont {P.}~\bibnamefont {Prelovšek}}, \bibinfo {editor}
  {\bibfnamefont {A.}~\bibnamefont {Ramšak}},\ and\ \bibinfo {editor}
  {\bibfnamefont {S.}~\bibnamefont {Sarkar}}}\ (\bibinfo  {publisher} {Springer
  Netherlands},\ \bibinfo {address} {Dordrecht},\ \bibinfo {year} {2001})\ p.\
  \bibinfo {pages} {273}\BibitemShut {NoStop}%
\bibitem [{\citenamefont {Auerbach}\ and\ \citenamefont
  {Bhattacharyya}(2024)}]{auerbach_quantum_2024}%
  \BibitemOpen
  \bibfield  {author} {\bibinfo {author} {\bibfnamefont {A.}~\bibnamefont
  {Auerbach}}\ and\ \bibinfo {author} {\bibfnamefont {S.}~\bibnamefont
  {Bhattacharyya}},\ }\bibfield  {title} {\bibinfo {title} {Quantum {Transport}
  {Theory} of {Strongly} {Correlated} {Matter}},\ }\href
  {https://doi.org/10.1016/j.physrep.2024.09.005} {\bibfield  {journal}
  {\bibinfo  {journal} {Physics Reports}\ }\textbf {\bibinfo {volume} {1091}},\
  \bibinfo {pages} {1} (\bibinfo {year} {2024})},\ \bibinfo {note}
  {arXiv:2406.02677 [cond-mat]}\BibitemShut {NoStop}%
\bibitem [{\citenamefont {Schulz}(1980)}]{schulz_cic2d}%
  \BibitemOpen
  \bibfield  {author} {\bibinfo {author} {\bibfnamefont {H.~J.}\ \bibnamefont
  {Schulz}},\ }\href@noop {} {\bibfield  {journal} {\bibinfo  {journal}
  {Physical Review B}\ }\textbf {\bibinfo {volume} {22}},\ \bibinfo {pages}
  {5274} (\bibinfo {year} {1980})}\BibitemShut {NoStop}%
\bibitem [{\citenamefont {{Bateman Manuscript
  Proyect}}(1953)}]{erdelyi_special_functions}%
  \BibitemOpen
  \bibfield  {author} {\bibinfo {author} {\bibnamefont {{Bateman Manuscript
  Proyect}}},\ }\bibinfo {title} {Higher transcendental functions}\ (\bibinfo
  {publisher} {McGraw-Hill Book Company Inc.},\ \bibinfo {address} {New York},\
  \bibinfo {year} {1953})\BibitemShut {NoStop}%
\bibitem [{\citenamefont {Olver}\ \emph {et~al.}(2010)\citenamefont {Olver},
  \citenamefont {Lozier}, \citenamefont {Boisvert},\ and\ \citenamefont
  {Clark}}]{olver_nist_2010}%
  \BibitemOpen
  \bibinfo {editor} {\bibfnamefont {F.}~\bibnamefont {Olver}}, \bibinfo
  {editor} {\bibfnamefont {D.}~\bibnamefont {Lozier}}, \bibinfo {editor}
  {\bibfnamefont {R.}~\bibnamefont {Boisvert}},\ and\ \bibinfo {editor}
  {\bibfnamefont {C.}~\bibnamefont {Clark}},\ eds.,\ \href
  {https://dlmf.nist.gov} {\emph {\bibinfo {title} {{NIST} handbook of
  mathematical functions}}}\ (\bibinfo  {publisher} {Cambridge University
  Press},\ \bibinfo {address} {Cambridge, UK},\ \bibinfo {year}
  {2010})\BibitemShut {NoStop}%
\bibitem [{\citenamefont {Shastry}\ \emph
  {et~al.}(1993{\natexlab{b}})\citenamefont {Shastry}, \citenamefont
  {Shraiman},\ and\ \citenamefont {Singh}}]{Shastry1993}%
  \BibitemOpen
  \bibfield  {author} {\bibinfo {author} {\bibfnamefont {B.~S.}\ \bibnamefont
  {Shastry}}, \bibinfo {author} {\bibfnamefont {B.~I.}\ \bibnamefont
  {Shraiman}},\ and\ \bibinfo {author} {\bibfnamefont {R.~R.}\ \bibnamefont
  {Singh}},\ }\bibfield  {title} {\bibinfo {title} {Faraday rotation and the
  hall constant in strongly correlated fermi systems},\ }\href@noop {}
  {\bibfield  {journal} {\bibinfo  {journal} {Physical review letters}\
  }\textbf {\bibinfo {volume} {70}},\ \bibinfo {pages} {2004} (\bibinfo {year}
  {1993}{\natexlab{b}})}\BibitemShut {NoStop}%
\bibitem [{\citenamefont {Lieb}\ and\ \citenamefont
  {Liniger}(1963)}]{lieb_bosons_1D}%
  \BibitemOpen
  \bibfield  {author} {\bibinfo {author} {\bibfnamefont {E.~H.}\ \bibnamefont
  {Lieb}}\ and\ \bibinfo {author} {\bibfnamefont {W.}~\bibnamefont {Liniger}},\
  }\href@noop {} {\bibfield  {journal} {\bibinfo  {journal} {Physical Review}\
  }\textbf {\bibinfo {volume} {130}},\ \bibinfo {pages} {1605} (\bibinfo {year}
  {1963})}\BibitemShut {NoStop}%
\bibitem [{\citenamefont {Amico}\ and\ \citenamefont
  {Korepin}(2004)}]{amico04_boson_integrable_review}%
  \BibitemOpen
  \bibfield  {author} {\bibinfo {author} {\bibfnamefont {L.}~\bibnamefont
  {Amico}}\ and\ \bibinfo {author} {\bibfnamefont {V.}~\bibnamefont
  {Korepin}},\ }\bibfield  {title} {\bibinfo {title} {Universality of the
  one-dimensional bose gas with delta interaction},\ }\href@noop {} {\bibfield
  {journal} {\bibinfo  {journal} {Annalen der Physik}\ }\textbf {\bibinfo
  {volume} {314}},\ \bibinfo {pages} {496} (\bibinfo {year}
  {2004})}\BibitemShut {NoStop}%
\bibitem [{\citenamefont {Luther}\ and\ \citenamefont
  {Emery}(1974)}]{luther_emery_backscattering}%
  \BibitemOpen
  \bibfield  {author} {\bibinfo {author} {\bibfnamefont {A.}~\bibnamefont
  {Luther}}\ and\ \bibinfo {author} {\bibfnamefont {V.~J.}\ \bibnamefont
  {Emery}},\ }\bibfield  {title} {\bibinfo {title} {Backward scattering in the
  one-oimensional electron gas},\ }\href@noop {} {\bibfield  {journal}
  {\bibinfo  {journal} {Physical Review Letters}\ }\textbf {\bibinfo {volume}
  {33}},\ \bibinfo {pages} {589} (\bibinfo {year} {1974})}\BibitemShut
  {NoStop}%
\bibitem [{\citenamefont {Landau}\ and\ \citenamefont
  {Lifshitz}(1962)}]{landau_mecaq}%
  \BibitemOpen
  \bibfield  {author} {\bibinfo {author} {\bibfnamefont {L.~D.}\ \bibnamefont
  {Landau}}\ and\ \bibinfo {author} {\bibfnamefont {E.~M.}\ \bibnamefont
  {Lifshitz}},\ }\href@noop {} {\emph {\bibinfo {title} {Quantum Mechanics :
  non-relativistic theory}}}\ (\bibinfo  {publisher} {perg},\ \bibinfo
  {address} {New York},\ \bibinfo {year} {1962})\BibitemShut {NoStop}%
\bibitem [{\citenamefont {Mahan}(1981)}]{mahan_book}%
  \BibitemOpen
  \bibfield  {author} {\bibinfo {author} {\bibfnamefont {G.~D.}\ \bibnamefont
  {Mahan}},\ }\href@noop {} {\emph {\bibinfo {title} {Many Particle Physics}}}\
  (\bibinfo  {publisher} {Plenum},\ \bibinfo {address} {New York},\ \bibinfo
  {year} {1981})\BibitemShut {NoStop}%
\bibitem [{\citenamefont {Essler}\ and\ \citenamefont
  {Konik}(2004)}]{essler04_condmat_exact_review}%
  \BibitemOpen
  \bibfield  {author} {\bibinfo {author} {\bibfnamefont {F.~H.}\ \bibnamefont
  {Essler}}\ and\ \bibinfo {author} {\bibfnamefont {R.~M.}\ \bibnamefont
  {Konik}},\ }\bibfield  {title} {\bibinfo {title} {Applications of massive
  integrable quantum field theories to problems in condensed matter physics}}
  (\bibinfo {year} {2004}),\ \bibinfo {note} {cond-mat/0412421}\BibitemShut
  {NoStop}%
\bibitem [{\citenamefont {Karowski}\ and\ \citenamefont
  {Wiesz}(1978)}]{karowski_ff}%
  \BibitemOpen
  \bibfield  {author} {\bibinfo {author} {\bibfnamefont {M.}~\bibnamefont
  {Karowski}}\ and\ \bibinfo {author} {\bibfnamefont {P.}~\bibnamefont
  {Wiesz}},\ }\bibfield  {title} {\bibinfo {title} {Exact form factors in
  (1+1)dimensional field theoretic models with soliton behavior},\ }\href@noop
  {} {\bibfield  {journal} {\bibinfo  {journal} {Nuclear Physics B}\ }\textbf
  {\bibinfo {volume} {139}},\ \bibinfo {pages} {455} (\bibinfo {year}
  {1978})}\BibitemShut {NoStop}%
\bibitem [{\citenamefont {Essler}\ and\ \citenamefont
  {Konik}(2009)}]{essler-konik_2009}%
  \BibitemOpen
  \bibfield  {author} {\bibinfo {author} {\bibfnamefont {F.~H.~L.}\
  \bibnamefont {Essler}}\ and\ \bibinfo {author} {\bibfnamefont {R.~M.}\
  \bibnamefont {Konik}},\ }\bibfield  {title} {\bibinfo {title}
  {Finite-temperature dynamical correlations in massive integrable quantum
  field theories},\ }\href {https://doi.org/10.1088/1742-5468/2009/09/P09018}
  {\bibfield  {journal} {\bibinfo  {journal} {Journal of Statistical Mechanics:
  Theory and Experiment}\ }\textbf {\bibinfo {volume} {2009}},\ \bibinfo
  {pages} {P09018} (\bibinfo {year} {2009})}\BibitemShut {NoStop}%
\bibitem [{\citenamefont {Schulz}(1994)}]{schulz_losalamos}%
  \BibitemOpen
  \bibfield  {author} {\bibinfo {author} {\bibfnamefont {H.~J.}\ \bibnamefont
  {Schulz}},\ }\bibinfo {title} {Strongly correlated electronic materials: The
  los alamos symposium 1993}\ (\bibinfo  {publisher} {Addison--Wesley},\
  \bibinfo {address} {Reading, Massachusetts},\ \bibinfo {year} {1994})\ p.\
  \bibinfo {pages} {187}\BibitemShut {NoStop}%
\bibitem [{\citenamefont {Buser}\ \emph {et~al.}(2021)\citenamefont {Buser},
  \citenamefont {Greschner}, \citenamefont {Schollw\"ock},\ and\ \citenamefont
  {Giamarchi}}]{buser_hall_ladder}%
  \BibitemOpen
  \bibfield  {author} {\bibinfo {author} {\bibfnamefont {M.}~\bibnamefont
  {Buser}}, \bibinfo {author} {\bibfnamefont {S.}~\bibnamefont {Greschner}},
  \bibinfo {author} {\bibfnamefont {U.}~\bibnamefont {Schollw\"ock}},\ and\
  \bibinfo {author} {\bibfnamefont {T.}~\bibnamefont {Giamarchi}},\ }\bibfield
  {title} {\bibinfo {title} {Probing the hall voltage in synthetic quantum
  systems},\ }\href {https://doi.org/10.1103/PhysRevLett.126.030501} {\bibfield
   {journal} {\bibinfo  {journal} {Phys. Rev. Lett.}\ }\textbf {\bibinfo
  {volume} {126}},\ \bibinfo {pages} {030501} (\bibinfo {year}
  {2021})}\BibitemShut {NoStop}%
\bibitem [{\citenamefont {Guan}\ \emph {et~al.}(2020)\citenamefont {Guan},
  \citenamefont {Feng}, \citenamefont {Xue}, \citenamefont {Chen},\ and\
  \citenamefont {Jia}}]{guan2020}%
  \BibitemOpen
  \bibfield  {author} {\bibinfo {author} {\bibfnamefont {X.}~\bibnamefont
  {Guan}}, \bibinfo {author} {\bibfnamefont {Y.}~\bibnamefont {Feng}}, \bibinfo
  {author} {\bibfnamefont {Z.-Y.}\ \bibnamefont {Xue}}, \bibinfo {author}
  {\bibfnamefont {G.}~\bibnamefont {Chen}},\ and\ \bibinfo {author}
  {\bibfnamefont {S.}~\bibnamefont {Jia}},\ }\bibfield  {title} {\bibinfo
  {title} {Synthetic gauge field and chiral physics on two-leg superconducting
  circuits},\ }\href {https://doi.org/10.1103/PhysRevA.102.032610} {\bibfield
  {journal} {\bibinfo  {journal} {Phys. Rev. A}\ }\textbf {\bibinfo {volume}
  {102}},\ \bibinfo {pages} {032610} (\bibinfo {year} {2020})}\BibitemShut
  {NoStop}%
\bibitem [{\citenamefont {Lecheminant}\ and\ \citenamefont
  {Nonne}(2012)}]{lecheminant_exotic_2012}%
  \BibitemOpen
  \bibfield  {author} {\bibinfo {author} {\bibfnamefont {P.}~\bibnamefont
  {Lecheminant}}\ and\ \bibinfo {author} {\bibfnamefont {H.}~\bibnamefont
  {Nonne}},\ }\bibfield  {title} {\bibinfo {title} {Exotic quantum criticality
  in one-dimensional coupled dipolar bosons tubes},\ }\href
  {https://doi.org/10.1103/PhysRevB.85.195121} {\bibfield  {journal} {\bibinfo
  {journal} {Physical Review B}\ }\textbf {\bibinfo {volume} {85}},\ \bibinfo
  {pages} {195121} (\bibinfo {year} {2012})}\BibitemShut {NoStop}%
\bibitem [{\citenamefont {Bertini}\ \emph {et~al.}(2021)\citenamefont
  {Bertini}, \citenamefont {Heidrich-Meisner}, \citenamefont {Karrasch},
  \citenamefont {Prosen}, \citenamefont {Steinigeweg},\ and\ \citenamefont
  {Žnidarič}}]{bertini-review_2021}%
  \BibitemOpen
  \bibfield  {author} {\bibinfo {author} {\bibfnamefont {B.}~\bibnamefont
  {Bertini}}, \bibinfo {author} {\bibfnamefont {F.}~\bibnamefont
  {Heidrich-Meisner}}, \bibinfo {author} {\bibfnamefont {C.}~\bibnamefont
  {Karrasch}}, \bibinfo {author} {\bibfnamefont {T.}~\bibnamefont {Prosen}},
  \bibinfo {author} {\bibfnamefont {R.}~\bibnamefont {Steinigeweg}},\ and\
  \bibinfo {author} {\bibfnamefont {M.}~\bibnamefont {Žnidarič}},\ }\bibfield
   {title} {\bibinfo {title} {Finite-temperature transport in one-dimensional
  quantum lattice models},\ }\href
  {https://doi.org/10.1103/RevModPhys.93.025003} {\bibfield  {journal}
  {\bibinfo  {journal} {Reviews of Modern Physics}\ }\textbf {\bibinfo {volume}
  {93}},\ \bibinfo {pages} {025003} (\bibinfo {year} {2021})}\BibitemShut
  {NoStop}%
\bibitem [{\citenamefont {Caldeira}\ and\ \citenamefont
  {Leggett}(1983)}]{caldeira_leggett}%
  \BibitemOpen
  \bibfield  {author} {\bibinfo {author} {\bibfnamefont {A.~O.}\ \bibnamefont
  {Caldeira}}\ and\ \bibinfo {author} {\bibfnamefont {A.~J.}\ \bibnamefont
  {Leggett}},\ }\href@noop {} {\bibfield  {journal} {\bibinfo  {journal}
  {Annalen der Physik}\ }\textbf {\bibinfo {volume} {149}},\ \bibinfo {pages}
  {374} (\bibinfo {year} {1983})}\BibitemShut {NoStop}%
\bibitem [{\citenamefont {Ambegaokar}\ \emph {et~al.}(1982)\citenamefont
  {Ambegaokar}, \citenamefont {Eckern},\ and\ \citenamefont
  {{Sch{\"o}n}}}]{ambegaokar_josephson_dissipation_short}%
  \BibitemOpen
  \bibfield  {author} {\bibinfo {author} {\bibfnamefont {V.}~\bibnamefont
  {Ambegaokar}}, \bibinfo {author} {\bibfnamefont {U.}~\bibnamefont {Eckern}},\
  and\ \bibinfo {author} {\bibfnamefont {G.}~\bibnamefont {{Sch{\"o}n}}},\
  }\bibfield  {title} {\bibinfo {title} {Quantum dynamics of tunneling between
  superconductors},\ }\href@noop {} {\bibfield  {journal} {\bibinfo  {journal}
  {Physical Review Letters}\ }\textbf {\bibinfo {volume} {48}},\ \bibinfo
  {pages} {1745} (\bibinfo {year} {1982})}\BibitemShut {NoStop}%
\bibitem [{\citenamefont {Majumdar}\ \emph {et~al.}(2023)\citenamefont
  {Majumdar}, \citenamefont {Foini}, \citenamefont {Giamarchi},\ and\
  \citenamefont {Rosso}}]{majumdar_bath2023}%
  \BibitemOpen
  \bibfield  {author} {\bibinfo {author} {\bibfnamefont {S.}~\bibnamefont
  {Majumdar}}, \bibinfo {author} {\bibfnamefont {L.}~\bibnamefont {Foini}},
  \bibinfo {author} {\bibfnamefont {T.}~\bibnamefont {Giamarchi}},\ and\
  \bibinfo {author} {\bibfnamefont {A.}~\bibnamefont {Rosso}},\ }\bibfield
  {title} {\bibinfo {title} {Bath-induced phase transition in a {Luttinger}
  liquid},\ }\href {https://doi.org/10.1103/PhysRevB.107.165113} {\bibfield
  {journal} {\bibinfo  {journal} {Physical Review B}\ }\textbf {\bibinfo
  {volume} {107}},\ \bibinfo {pages} {165113} (\bibinfo {year} {2023})},\
  \bibinfo {note} {arXiv:2210.01590 [cond-mat]}\BibitemShut {NoStop}%
\bibitem [{\citenamefont {Guo}\ and\ \citenamefont
  {Poletti}(2017)}]{guo-lindblad_2017}%
  \BibitemOpen
  \bibfield  {author} {\bibinfo {author} {\bibfnamefont {C.}~\bibnamefont
  {Guo}}\ and\ \bibinfo {author} {\bibfnamefont {D.}~\bibnamefont {Poletti}},\
  }\bibfield  {title} {\bibinfo {title} {Dissipatively driven strongly
  interacting bosons in a gauge field},\ }\href@noop {} {\bibfield  {journal}
  {\bibinfo  {journal} {Phys. Rev. B}\ }\textbf {\bibinfo {volume} {96}},\
  \bibinfo {pages} {165409} (\bibinfo {year} {2017})},\ \bibinfo {note}
  {arXiv:1705.07633}\BibitemShut {NoStop}%
\bibitem [{\citenamefont {Zhou}\ \emph
  {et~al.}(2023{\natexlab{b}})\citenamefont {Zhou}, \citenamefont {Cappellini},
  \citenamefont {Tusi}, \citenamefont {Franchi}, \citenamefont {Parravicini},
  \citenamefont {Repellin}, \citenamefont {Greschner}, \citenamefont
  {Inguscio}, \citenamefont {Giamarchi}, \citenamefont {Filippone} \emph
  {et~al.}}]{zhou_2022}%
  \BibitemOpen
  \bibfield  {author} {\bibinfo {author} {\bibfnamefont {T.-W.}\ \bibnamefont
  {Zhou}}, \bibinfo {author} {\bibfnamefont {G.}~\bibnamefont {Cappellini}},
  \bibinfo {author} {\bibfnamefont {D.}~\bibnamefont {Tusi}}, \bibinfo {author}
  {\bibfnamefont {L.}~\bibnamefont {Franchi}}, \bibinfo {author} {\bibfnamefont
  {J.}~\bibnamefont {Parravicini}}, \bibinfo {author} {\bibfnamefont
  {C.}~\bibnamefont {Repellin}}, \bibinfo {author} {\bibfnamefont
  {S.}~\bibnamefont {Greschner}}, \bibinfo {author} {\bibfnamefont
  {M.}~\bibnamefont {Inguscio}}, \bibinfo {author} {\bibfnamefont
  {T.}~\bibnamefont {Giamarchi}}, \bibinfo {author} {\bibfnamefont
  {M.}~\bibnamefont {Filippone}}, \emph {et~al.},\ }\bibfield  {title}
  {\bibinfo {title} {Observation of universal hall response in strongly
  interacting fermions},\ }\bibfield  {journal} {\bibinfo  {journal} {Science}\
  }\textbf {\bibinfo {volume} {381}},\ \href
  {https://doi.org/10.1126/science.add1969} {10.1126/science.add1969} (\bibinfo
  {year} {2023}{\natexlab{b}}),\ \bibinfo {note} {arXiv:2205.13567}\BibitemShut
  {NoStop}%
\bibitem [{\citenamefont {Zhou}\ \emph {et~al.}(2025)\citenamefont {Zhou},
  \citenamefont {Beller}, \citenamefont {Masini}, \citenamefont {Parravicini},
  \citenamefont {Cappellini}, \citenamefont {Repellin}, \citenamefont
  {Giamarchi}, \citenamefont {Catani}, \citenamefont {Filippone},\ and\
  \citenamefont {Fallani}}]{zhou_2025}%
  \BibitemOpen
  \bibfield  {author} {\bibinfo {author} {\bibfnamefont {T.-W.}\ \bibnamefont
  {Zhou}}, \bibinfo {author} {\bibfnamefont {T.}~\bibnamefont {Beller}},
  \bibinfo {author} {\bibfnamefont {G.}~\bibnamefont {Masini}}, \bibinfo
  {author} {\bibfnamefont {J.}~\bibnamefont {Parravicini}}, \bibinfo {author}
  {\bibfnamefont {G.}~\bibnamefont {Cappellini}}, \bibinfo {author}
  {\bibfnamefont {C.}~\bibnamefont {Repellin}}, \bibinfo {author}
  {\bibfnamefont {T.}~\bibnamefont {Giamarchi}}, \bibinfo {author}
  {\bibfnamefont {J.}~\bibnamefont {Catani}}, \bibinfo {author} {\bibfnamefont
  {M.}~\bibnamefont {Filippone}},\ and\ \bibinfo {author} {\bibfnamefont
  {L.}~\bibnamefont {Fallani}},\ }\bibfield  {title} {\bibinfo {title}
  {Measuring {Hall} voltage and {Hall} resistance in an atom-based quantum
  simulator},\ }\href {https://doi.org/10.1038/s41467-025-65083-6} {\bibfield
  {journal} {\bibinfo  {journal} {Nature Communications}\ }\textbf {\bibinfo
  {volume} {16}},\ \bibinfo {pages} {10247} (\bibinfo {year} {2025})},\
  \bibinfo {note} {arXiv:2411.09744 [cond-mat]}\BibitemShut {NoStop}%
\end{thebibliography}
%

\end{document}